\documentstyle[amstex,epic,aps,epsfig,preprint]{revtex}

\begin{document}

\renewcommand{\theequation}{\arabic{section}.\arabic{equation}}

\setlength{\unitlength}{.5mm}

\author{ A.N. Morozov$^{1,2}$ \and J.G.E.M. Fraaije$^2$} 

\address{$^1$Faculty of Mathematics and Natural Sciences, University of
Groningen, Nijenborg 4, 9747 AG Groningen, The Netherlands, email: a.n.morozov@@chem.rug.nl}

\address{$^2$Soft Condensed Matter group, LIC, Leiden University,
PO Box 9502, 2300 RA Leiden, The Netherlands}

\title{Phase behaviour of block copolymer melts with arbitrary architecture}

\maketitle 

\begin{abstract}
The Leibler theory [L. Leibler, Macromolecules {\bf{13}} 1602 (1980)] for microphase separation in AB block
copolymer melts is generalized for systems with arbitrary topology of molecules. A diagrammatic technique  for
calculation of the monomeric correlation functions is developed. The free energies of various mesophases
are calculated within the second-harmonic approximation. Model highly-branched tree-like structures
are considered as an example and their phase diagrams are obtained. The topology of molecules is found to
influence the spinodal temperature and asymmetry of the phase diagrams, but not the types of phases and their
order. We suggest that all model AB block-copolymer systems will exhibit the typical phase behaviour.
\end{abstract}

\newpage

\section{Introduction}

One of the biggest challenges for theoretical polymer science is to predict the phase behaviour
of a polymer system with given molecular structure. The required theory should explicitely contain molecular parameters and be 
able to predict thermodynamically stable phases (whatever they are) and temperatures of transitions among them. Apart
from the theoretical interest this problem is of great industrial importance. Since the mechanical properties of the phases are 
studied quite well it is useful to predict the conditions at which these phases are stable. In the absence of such
predictions the experimental search could take even years. In the present situation of a rapidly increasing number of novel materials
even a crude theoretical model could help a great deal in direct applications.

Among the systems of interest are solutions of homopolymers, block copolymer melts, blends and so on. In this article we focus our 
attention on the A-B block copolymer melts. Such systems can only differ in the total content of one component, the length of 
molecules and the way the A and B blocks are arranged in one molecule. As a function of these parameters and temperature, the following
phases were observed experimentally: the lamellar phase (LAM), the hexagonal phase (HEX), the body-centered cubic phase 
(BCC) and the bicontinuous gyroid phase (GYR) \cite{Bates:1990B,hamley:book}. However, the structure of these phases strongly depends on 
temperature. Just below the order-disorder
transition temperature the chains are slightly stretched and the mesophases consist of periodic patterns with only a small 
increase in the content of one component. This situation is called the weak-segregation regime. On contrary, the strong-segregation 
regime (very low temperatures) is characterized by domains which are almost pure in one component. Both regimes are limiting cases 
and therefore can be studied analytically. In the intermediate regime there is no small parameter and only numerical investigations 
are possible \cite{matsen97}

The first weak-segregation theory of AB melts is by Leibler \cite{Leibler:1980} who considered a model Gaussian diblock copolymer. He was able to
estimate the order-disorder transition temperature and predict the following sequence of transitions: DIS$\rightarrow$BCC
$\rightarrow$HEX$\rightarrow$LAM with decrease of temperature. Later, after the GYR phase was observed experimentally \cite{gyr_exp1,gyr_exp2}, Milner and Olmsted \cite{milner97} used the second-harmonic approximation and predicted the GYR phase to be stable in between the HEX and LAM phases.
The same conclusion was drawn in \cite{matsen_bates,matsen97} and in the Landau-Brazovskii theory of Hamley and Podneks \cite{podneks96,hamley97}.

The method developed by Leibler was applied to other topologies of block copolymer molecules.
Thus, Mayes and de la Cruz considered an ABA triblock molecule \cite{mayes89}, de la Cruz and Sanchez studied star- and simple graft-copolymers \cite{Cruz:1986};
Dobrynin and Erukhimovich developed a computer-aided tool to calculate phase diagrams for various topologies \cite{Dobrynin:1993}. However,
in all these studies the authors started with a fixed molecular topology. To consider another topology one needed to go through the
whole calculation procedure again, from the very beginning. In this work we show how the Leibler approach can be used for an 
arbitrary topology of molecules. Section I repeats some main points of Leibler's paper \cite{Leibler:1980} and introduces a diagrammatic technique
that helps to deal with molecules of  arbitrary topology. The first step in this direction was made by Read \cite{Read:1998}, who invented a
way to calculate structure factors of polymers of arbitrary complexity. We go further and  show how to calculate higher vertex functions
for such systems. Section II describes the way to calculate the free energies of 
different mesophases and discusses some of the approximations. We consider the face-centered cubic phase (FCC), the ordered 
bicontinuous double-diamond phase, the simple cubic (SC) and the square lattice (SL) phases in addition to the classical phases. 
In the Results and Discussion we apply the developed technique to a model dendrimer system which has recently attracted 
wide experimental \cite{Hadji:review} and theoretical \cite{dend:theory} attention. The phase diagrams of various dendrimers are presented and their features 
are discussed. The results for many more systems that were not included in this publication are presented as supplementary materials
on the Internet: {\it http://rugmd4.chem.rug.nl/\~{}morozov/research.html}. Finally, we summarize our work in the Conclusion and propose
some possible developments of the method.

\section{Mean-field theory}
\setcounter{equation}{0} 

A mean-field theory of incompressible AB block copolymer melts was first formulated
by Leibler \cite{Leibler:1980}. In his approach a local deviation of the density of A component from
its average value plays the role of an order parameter $\psi({\mathbf{r}})$. He found that the
 coefficients in the phenomenological free energy expansion

\begin{eqnarray}
\label{Hamiltonian}
F=F_0&+&\frac{1}{2}\int_{q_1}\int_{q_2}\Gamma_2({\mathbf{q}}_1,{\mathbf{q}}_2)\psi({\mathbf{q}}_1)\psi({\mathbf{q}}_2)+
\frac{1}{3!}\int_{q_1}\int_{q_2}\int_{q_3}\Gamma_3({\mathbf{q}}_1,{\mathbf{q}}_2,{\mathbf{q}}_3)\psi({\mathbf{q}}_1)\psi({\mathbf{q}}_2)\psi({\mathbf{q}}_3) \nonumber \\
&+&\frac{1}{4!}\int_{q_1}\int_{q_2}\int_{q_3}\int_{q_4}\Gamma_4({\mathbf{q}}_1,{\mathbf{q}}_2,{\mathbf{q}}_3,{\mathbf{q}}_4)\psi({\mathbf{q}}_1)\psi({\mathbf{q}}_2)\psi({\mathbf{q}}_3)\psi({\mathbf{q}}_4)+\cdots  
\end{eqnarray}
can be connected with microscopic properties of the system. This possibility is based on the assumption that
in a melt a single chain behaves as almost ideal. This constitutes the so-called random phase approximation (RPA)
which expresses the monomer density correlation functions in terms of the correlation functions of a Gaussian
chain. For the vertex functions in (\ref{Hamiltonian}) it gives \cite{Leibler:1980}:

\begin{eqnarray}
\label{vertex}
\Gamma_2({\mathbf{q}}_1,{\mathbf{q}}_2)&=&\delta({\mathbf{q}}_1 +{\mathbf{q}}_2 )
	\left[\frac{S_{AA}({\mathbf{q}}_1)+S_{BB}({\mathbf{q}}_1)+2S_{AB}({\mathbf{q}}_1)}{S_{AA}({\mathbf{q}}_1)S_{BB}({\mathbf{q}}_1)-S_{AB}^2({\mathbf{q}}_1)}-2\chi\right] \nonumber \\
\Gamma_3({\mathbf{q}}_1,{\mathbf{q}}_2,{\mathbf{q}}_3)&=&-G^{(3)}_{ijk}({\mathbf{q}}_1,{\mathbf{q}}_2,{\mathbf{q}}_3)
	\left[S^{-1}_{iA}({\mathbf{q}}_1)-S^{-1}_{iB}({\mathbf{q}}_1)\right]  \nonumber \\
	& &\times\left[S^{-1}_{jA}({\mathbf{q}}_2)-S^{-1}_{jB}({\mathbf{q}}_2)\right]
	\left[S^{-1}_{kA}({\mathbf{q}}_3)-S^{-1}_{kB}({\mathbf{q}}_3)\right] \\
\Gamma_4({\mathbf{q}}_1,{\mathbf{q}}_2,{\mathbf{q}}_3,{\mathbf{q}}_4)&=&\gamma_{ijkl}({\mathbf{q}}_1,{\mathbf{q}}_2,{\mathbf{q}}_3,{\mathbf{q}}_4) \left[S^{-1}_{iA}({\mathbf{q}}_1)-S^{-1}_{iB}({\mathbf{q}}_1)\right]
	\left[S^{-1}_{jA}({\mathbf{q}}_2)-S^{-1}_{jB}({\mathbf{q}}_2)\right] \nonumber \\
	& &\times\left[S^{-1}_{kA}({\mathbf{q}}_3)-S^{-1}_{kB}({\mathbf{q}}_3)\right]
	\left[S^{-1}_{lA}({\mathbf{q}}_4)-S^{-1}_{lB}({\mathbf{q}}_4)\right] \nonumber 
\end{eqnarray}
where
\begin{eqnarray}
\label{small_gamma}
& &\gamma_{ijkl}({\mathbf{q}}_1,{\mathbf{q}}_2,{\mathbf{q}}_3,{\mathbf{q}}_4)=-G^{(4)}_{ijkl}({\mathbf{q}}_1,{\mathbf{q}}_2,{\mathbf{q}}_3,{\mathbf{q}}_4) \nonumber \\
&+& \int_{p}S^{-1}_{mn}({\mathbf{p}})\left[ G^{(3)}_{ijm}({\mathbf{q}}_1,{\mathbf{q}}_2,{\mathbf{p}})G^{(3)}_{nkl}(-{\mathbf{p}},{\mathbf{q}}_3,{\mathbf{q}}_4)
+G^{(3)}_{ikm}({\mathbf{q}}_1,{\mathbf{q}}_3,{\mathbf{p}})G^{(3)}_{njl}(-{\mathbf{p}},{\mathbf{q}}_2,{\mathbf{q}}_4)\right. \\
&+&\left.G^{(3)}_{ilm}({\mathbf{q}}_1,{\mathbf{q}}_4,{\mathbf{p}})G^{(3)}_{nkj}(-{\mathbf{p}},{\mathbf{q}}_3,{\mathbf{q}}_2)\right] \nonumber 
\end{eqnarray}
The summation over repeating indices is assumed. The second, the third and the fourth order correlation functions of the Gaussian chain are denoted by $S_{ij}$, $G^{(3)}_{ijk}$ and $G^{(4)}_{ijkl}$, respectively ($i,j,k,l=A,B$); $S^{-1}_{ij}$ are the elements of the inverse matrix
$||S_{ij}||^{-1}$. The  Flory-Huggins parameter describing the strength of interaction between A and B monomers is denoted by $\chi$. 

A general method of calculating the Gaussian correlation functions was developed in Appendix B of \cite{Leibler:1980}. It is based on the fact
that the Gaussian distribution is fully determined by its first two moments. As a result the probability to find several monomers at
given positions on the chain is expressed as a product of the 2-point correlation functions of the consequently neighbouring
monomers along the chain. For example,

\begin{picture}(100,50)
\linethickness{.5pt}
\put(50,20){\line(1,0){80}}
\put(70,20){\circle*{3}}
\put(90,20){\circle*{3}}
\put(110,20){\circle*{3}}
\put(66,23){\makebox(10,10)[c]{${\mathbf{r}}_1$}}
\put(86,23){\makebox(10,10)[c]{${\mathbf{r}}_2$}}
\put(106,23){\makebox(10,10)[c]{${\mathbf{r}}_3$}}
\put(140,15){\makebox(10,10)[l]{$=$ $P({\mathbf{r}}_1,{\mathbf{r}}_2)P({\mathbf{r}}_2,{\mathbf{r}}_3)$}}
\end{picture}

The correlation functions $S_{ij}$, $G^{(3)}_{ijk}$ and $G^{(4)}_{ijkl}$ are calculated by averaging the Fourier transforms
of expressions like one in the previous example over all possible positions of monomers along the chain. At this point one needs
to introduce some topology of the polymer molecule in order to perform such an averaging. In \cite{Leibler:1980} this was done for a diblock
molecule. A triblock molecule was considered in \cite{mayes89}. More sophisticated architectures were studied in \cite{Dobrynin:1993}. We generalize
this procedure for arbitrary topology of a polymer molecule.

In the following we use the graphical method formulated in Appendix A. A correlation function of a given order is calculated by
drawing all possible (but topologically distinct) diagrams of this order. Thus,

\begin{picture}(100,50)
\linethickness{.5pt}
\put(80,15){\makebox(10,10)[l]{$S_{ij}$ $=$}}
\put(110,20){\line(1,0){40}}
\put(120,20){\circle*{3}}
\put(140,20){\circle*{3}}
\put(155,15){\makebox(10,10)[c]{$+$}}
\put(170,15){\line(1,0){40}}
\put(170,25){\line(1,0){40}}
\put(190,15){\circle*{3}}
\put(190,25){\circle*{3}}
\end{picture}

\begin{picture}(100,50)
\linethickness{.5pt}
\put(50,15){\makebox(10,10)[l]{$G^{(3)}_{ijk}$ $=$}}
\put(80,20){\line(1,0){40}}
\put(90,20){\circle*{3}}
\put(100,20){\circle*{3}}
\put(110,20){\circle*{3}}
\put(125,15){\makebox(10,10)[c]{$+$}}
\put(140,15){\line(1,0){40}}
\put(140,25){\line(1,0){40}}
\put(153,15){\circle*{3}}
\put(166,15){\circle*{3}}
\put(160,25){\circle*{3}}
\put(185,15){\makebox(10,10)[c]{$+$}}
\put(200,20){\line(1,0){40}}
\put(200,28){\line(1,0){40}}
\put(200,12){\line(1,0){40}}
\put(220,20){\circle*{3}}
\put(220,12){\circle*{3}}
\put(220,28){\circle*{3}}
\end{picture}

\begin{picture}(100,50)
\linethickness{.5pt}
\put(0,15){\makebox(10,10)[l]{$G^{(4)}_{ijkl}$ $=$}}
\put(30,20){\line(1,0){40}}
\put(38,20){\circle*{3}}
\put(46,20){\circle*{3}}
\put(54,20){\circle*{3}}
\put(62,20){\circle*{3}}
\put(75,15){\makebox(10,10)[c]{$+$}}
\put(90,15){\line(1,0){40}}
\put(90,25){\line(1,0){40}}
\put(100,15){\circle*{3}}
\put(110,15){\circle*{3}}
\put(120,15){\circle*{3}}
\put(110,25){\circle*{3}}
\put(135,15){\makebox(10,10)[c]{$+$}}
\put(150,15){\line(1,0){40}}
\put(150,25){\line(1,0){40}}
\put(163,15){\circle*{3}}
\put(176,15){\circle*{3}}
\put(163,25){\circle*{3}}
\put(176,25){\circle*{3}}
\put(195,15){\makebox(10,10)[c]{$+$}}
\put(210,20){\line(1,0){40}}
\put(210,28){\line(1,0){40}}
\put(210,12){\line(1,0){40}}
\put(223,12){\circle*{3}}
\put(236,12){\circle*{3}}
\put(230,20){\circle*{3}}
\put(230,28){\circle*{3}}
\put(255,15){\makebox(10,10)[c]{$+$}}
\put(270,16){\line(1,0){40}}
\put(270,8){\line(1,0){40}}
\put(270,24){\line(1,0){40}}
\put(270,32){\line(1,0){40}}
\put(290,16){\circle*{3}}
\put(290,8){\circle*{3}}
\put(290,24){\circle*{3}}
\put(290,32){\circle*{3}}
\end{picture}

This is just a schematic representation which shows that an $n$-th order diagram
can contain up to $n$ blocks. The corresponding contributions to the correlation
function are obtained by connecting these blocks in all possible ways. The topologically
different connections are presented in Appendix A. The contributions to the correlation
functions are calculated in Appendix B.

We want to note that {\it not all} diagrams contribute to the correlation functions of the
system with given topology. One restriction is connected with the type of the correlation
function we want to calculate. For example, the equation for $S_{AB}$ only contains diagrams
with the first label placed on some A block and the second label placed on a B block. Therefore
the first diagram in the general equation for $S_{ij}$ does not contribute to $S_{AB}$. The other
factor reducing the number of diagrams is a molecular architecture. Thus, for a linear multiblock
molecules all 'star' and 'fork' diagrams are impossible and should not be considered.

The final expressions for the correlation functions are obtained in the following way. Suppose we want
to calculate $G_{i_1,i_2,\dots}$. To do this we need to place the first label on any block of the type
$i_1$, the second label on any block of the type $i_2$ and so on. The resulting diagram will be one
of the diagrams calculated in Appendix B. The corresponding expression is written down. Then
we remove all labels and place them again avoiding previous configurations. This procedure is repeated 
until we covered all possible arrangements of the labels. The obtained expression is the equation for the
required $G_{i_1,i_2,\dots}$.

Once we have calculated the correlation functions for a given system, its free energy can be obtained with
the help of (\ref{Hamiltonian}-\ref{small_gamma}). In the next section this Landau-Ginzburg free energy 
will be a starting point for our analysis
of the phase transitions.

It is obvious that the formulated approach is general and within this framework all systems can be considered.

\section{Calculation of phase diagram}
\setcounter{equation}{0} 

In this section we show how to calculate a phase diagram of a system once its free energy functional
(\ref{Hamiltonian}) is known. Apart from some technical details this was developed elsewhere \cite{landau,Leibler:1980,marques90,milner97,Cruz:1986,Swift:1992,Ohta:1986,Lescanec:1993,Melenkevitz:1991,podneks96,hamley97}.

In the beginning we need to determine the order-disorder transition (ODT) temperature. The disordered phase can be described
by the Landau-Ginzburg Hamiltonian with only a quadratic term left \cite{landau,LL:stat}. The probability to create a spatial density fluctuation
$\psi({\mathbf{q}})$ is given by
\begin{equation}
P[\psi]\sim \exp\left[-\frac{1}{2}\int_{q}\Gamma_2(q)\psi({\mathbf{q}})\psi(-{\mathbf{q}}) \right]
\end{equation}
The typical life-time of such a fluctuation is $\displaystyle \tau\sim \frac{1}{\Gamma_2(q)}$. At some $q_*$, temperature 
(or Flory-Huggins parameter) and total content of A monomers in the molecule $\displaystyle \left(f=\frac{N_A}{N}\right)$,  $\Gamma_2(q_*)$
reaches zero. The probability to create such a mode tends to unity, while its life-time $\tau_{q_*}$ diverges.
It means that at this temperature the disordered phase becomes absolutely unstable with respect to the appearance of an ordered
mesophase with typical domain size $\sim q_*^{-1}$. Therefore, the set of equations:
\begin{eqnarray}
\Gamma_2(q_*,(\chi N)_s)&&=0 \\
\left.\frac{\partial \Gamma_2(q,(\chi N)_s)}{\partial q}\right|_{q_*}=0 \qquad,&&\qquad \left.\frac{\partial^2 \Gamma_2(q,(\chi N)_s)}{\partial q ^2}\right|_{q_*}>0 \nonumber
\end{eqnarray}
determines the spinodal temperature (which is very close to the ODT value \cite{Leibler:1980}) and the typical lengthscale $q_*$.

As the next step we choose a set of mesophases which will possibly enter the phase diagram. Then, for
each mesophase we represent the order parameter field $\psi({\mathbf{q}})$ in such a way that it possesses the
symmetry of the mesophase. This representation necessarily contains some unknown parameters. Next 
$\psi({\mathbf{q}})$ should be plugged in the eq.(\ref{Hamiltonian}) and the resulting expression should be minimized with
respect to the parameters. This procedure gives the free energies of different mesophases as functions of
temperature and composition. The molecular properties enter via the coefficients in eq.(\ref{Hamiltonian}). For the given temperature (below the ODT) 
and composition $f$ the mesophase with the lowest free energy will be the stable one. The phase diagram is drawn by varying 
temperature and $f$ over some ranges.

The way to ascribe a given symmetry to the order parameter is to expand it in the full orthonormal set of
functions with the given symmetry. The most popular choice is to use an expansion in plane waves:
\begin{equation}
\label{expansion}
\psi({\mathbf{q}})=\sum_{i}A_{i}\left(e^{i\phi_i}\delta_{{\mathbf{q}},{\mathbf{k}}_i}+e^{-i\phi_i}\delta_{{\mathbf{q}},-{\mathbf{k}}_i}\right)
\end{equation}
where $\{{\mathbf{k}}_i\}$ is a set of vectors defining a lattice of the given symmetry, $\phi_i$ and $A_i$ are the arbitrary
phases and amplitudes; $\delta_{i,j}$ denotes the Kronecker delta symbol. If the rotational symmetry in the system is not broken
we can rewrite (\ref{expansion}) in the following way:
\begin{equation}
\label{mod_exp}
\psi({\mathbf{q}})=\sum_{i}a_i\sum_{j=1}^{m_i}\left(e^{i\phi_j^{(i)}}\delta_{{\mathbf{q}},{\mathbf{k}}_j^{(i)}}+e^{-i\phi_j^{(i)}}\delta_{{\mathbf{q}},-{\mathbf{k}}_j^{(i)}}\right)
\end{equation}
where the first sum runs over the Brillouin zones or coordination spheres in the reciprocal space; $m_i$ is the number of vectors in the $i$-th zone. In practice we consider only the first few terms in the series (\ref{mod_exp}). The mathematical reason for that is to
simplify the minimization procedure. From the physical point of view, immediately below the transition temperature the appearing
mesophase is not well developed and we do not need too many details in the representation (\ref{mod_exp}). (As a rule, the length of 
the $q$-vectors increases with the number of the zone. Therefore, the more terms are left in (\ref{mod_exp}), the more detailed density 
profile we get.) The next question is how many terms we should leave in (\ref{mod_exp}). Unfortunately, there is no quantitative criterion where to cut the density expansion. In the past, several possibilities were considered. In the classical work \cite{Leibler:1980} the expansion 
(\ref{mod_exp}) was restricted to $i=1$. The so-called {\it second-harmonic approximation} ($i=2$) \cite{marques90} was 
formulated in \cite{hamley94,milner97,gurovich,Roan:1998,Swift:1992} in order to study complicated bicontinuous phases. Up to several hundred harmonics were used in the 
numerical method of Matsen \cite{matsen_bates} (however, the authors studied the intermediate/strong segregation regimes and, therefore, 
our discussion is not directly applicable to their model). Comparison of their results shows that higher harmonics not only result in
some non-significant shift of the phase boundaries, but also stabilize or destabilize several morphologies. In such a situation
the number of harmonics is crucial.

In this paper we adopt the criterion derived in \cite{hamley97,podneks96}. Instead of the mean-field theory Hamley and Podneks studied system
with composition fluctuations. In their model the cut-off value for $q_n$ (the $q$-vector from the $n$th shell) appears naturally as
an adaptation of the Ginzburg criterion:
\begin{equation}
\label{criterion}
\left[\left(\frac{q_n}{q_*}\right)^2-1 \right]^2 \ll 1
\end{equation} 
We apply this criterion  to our mean-field theory in order to keep 
uniformity with the fluctuation approach (a similar criterion was used in \cite{Holyst:1999}). 
There are several ways to support our choice. One type of argument is connected with the applicability region for our theory.
In the RPA there is essentially one lengthscale associated with the gyration radius of molecules. It is intrinsically connected
with the typical wavevector $q_*$. Therefore, the representation of the density profile by higher harmonics with $q\gg q_*$
is in strong contradiction with the used theory. (A similar idea was formulated in a different way in \cite{binder}.) One can also speculate that near the 
spinodal the inequality (\ref{criterion}) will always hold 
in the proper temperature range. The questions of applicability of (\ref{criterion}) and its influence on the phase diagram
will be addressed in future studies.

The cut-off (\ref{criterion}) implies that only one harmonic should be used for LAM, HEX, BCC, SC and SL phases, while FCC, GYR 
and OBDD phases are represented by two-harmonic density profile. The basis vectors for LAM, HEX, BCC, SC and SL phases, as well
as the first-shell vectors for FCC and OBDD (which coincides with the BCC first shell), are presented in \cite{Leibler:1980,marques90}.
For the rest we use (more detailed information can be found in \cite{Cryst.Tables}):

$$
\begin{array}{cccc}
\text{FCC:}\qquad & k_1^{(2)}=\frac{2q_*}{\sqrt{3}}(1,0,0)\quad &  k_2^{(2)}=\frac{2q_*}{\sqrt{3}}(0,1,0)\quad & 
k_3^{(2)}=\frac{2q_*}{\sqrt{3}}(0,0,1) \\
\text{OBDD:}\qquad & k_1^{(2)}=\frac{q_*}{\sqrt{2}}(1,1,1)\quad & k_2^{(2)}=\frac{q_*}{\sqrt{2}}(-1,-1,1)\quad & 
k_3^{(2)}=\frac{q_*}{\sqrt{2}}(-1,1,-1)\quad \\
& & k_4^{(2)}=\frac{q_*}{\sqrt{2}}(1,-1,-1)  & \\

\text{GYR:}\qquad & k_1^{(1)}=\frac{q_*}{\sqrt{6}}(1,-1,-2)\quad & k_2^{(1)}=\frac{q_*}{\sqrt{6}}(-1,-1,-2)\quad & 
k_3^{(1)}=\frac{q_*}{\sqrt{6}}(1,2,-1)\quad \\
& k_4^{(1)}=\frac{q_*}{\sqrt{6}}(-1,-2,-1,)\quad & k_5^{(1)}=\frac{q_*}{\sqrt{6}}(-2,1,-1)\quad & 
k_6^{(1)}=\frac{q_*}{\sqrt{6}}(-2,-1,-1)\quad \\
& k_7^{(1)}=\frac{q_*}{\sqrt{6}}(1,1,-2)\quad & k_8^{(1)}=\frac{q_*}{\sqrt{6}}(-1,1,-2)\quad & 
k_9^{(1)}=\frac{q_*}{\sqrt{6}}(-1,2,-1)\quad \\
& k_{10}^{(1)}=\frac{q_*}{\sqrt{6}}(1,-2,-1)\quad & k_{11}^{(1)}=\frac{q_*}{\sqrt{6}}(2,1,-1)\quad & 
k_{12}^{(1)}=\frac{q_*}{\sqrt{6}}(2,-1,-1)\quad  \\
& & & \\
& k_1^{(2)}=\frac{2q_*}{\sqrt{6}}(0,1,1)\quad & k_2^{(2)}=\frac{2q_*}{\sqrt{6}}(0,1,-1)\quad & 
k_3^{(2)}=\frac{2q_*}{\sqrt{6}}(1,1,0)\quad \\
& k_{4}^{(2)}=\frac{2q_*}{\sqrt{6}}(1,-1,0)\quad & k_{5}^{(2)}=\frac{2q_*}{\sqrt{6}}(1,0,1)\quad & 
k_{6}^{(2)}=\frac{2q_*}{\sqrt{6}}(1,0,-1)\quad 
\end{array}
$$

Once we have defined the density profile (\ref{mod_exp}), we plug it into the equation for the free energy
(\ref{Hamiltonian}). The integration is trivial since $\psi({\mathbf{q}})$ is a sum of delta-functions. The resulting
expression contains terms proportional to the vertex functions $\Gamma_2,\Gamma_3$ and $\Gamma_4$ which arguments are
all possible combinations of two, three and four $k_j^{(i)}$ that add up to zero. For the phases represented by
the two-harmonic density profile some of these terms will be the functions of the wave-vectors from the different shells.
However, while calculating the vertex functions (Section 2 and Appendix A and B) we assumed everywhere that all wave-vectors
have the same  modulus $q_*$. Correspondingly, we obtained the $m$-th order vertex function in the form 
$\Gamma_m(q_* {\mathbf{n}}_1,\dots,q_* {\mathbf{n}}_m)=\Gamma_m(x_*,h^{(m)})$ 
($\displaystyle {\mathbf{n}}_i=\frac{{\mathbf{q}}_i}{q_i}$; $h^{(m)}$ and $x$ are defined in the Appendix A). This
function cannot be directly used for the arguments with the different moduli. However, the following trick helps
to avoid this problem. 

The criterion (\ref{criterion}) implies that $q_j=q_*+\Delta_j$, where $\Delta_j\ll q_*$. Therefore, to the first order
in $\Delta$ we can approximate the $m$-th order vertex function of q-s with the different length by
$$
\Gamma_m({\mathbf{q}}_1,\dots,{\mathbf{q}}_m)\approx\Gamma_m(q_* {\mathbf{n}}_1,\dots,q_* {\mathbf{n}}_m)+
\sum_{\alpha=1}^{m}\kappa_\alpha \Delta_\alpha \quad,
$$
where
$$
\left.\kappa_\alpha=\frac{\partial \Gamma_m(q_* {\mathbf{n}}_1,\dots,{\mathbf{q}}_\alpha,\dots,q_* {\mathbf{n}}_m)}{\partial q_\alpha}
\right|_{q_\alpha=q_*}
$$
Any vertex function is a symmetric function of its arguments. Therefore, $\kappa_\alpha=\kappa$ is independent on $\alpha$. Let us now
put all $\Delta_\alpha=\Delta$. This gives
\begin{equation}
\label{real_gamma}
\Gamma_m({\mathbf{q}}_1,\dots,{\mathbf{q}}_m)\approx\Gamma_m(q_* {\mathbf{n}}_1,\dots,q_* {\mathbf{n}}_m)+
m \kappa \Delta
\end{equation}
But this expression should be identical to:
\begin{equation}
\label{my_gamma}
\left.\Gamma_m(x_*+\delta x, h^{(m)})\approx \Gamma_m(x_*, h^{(m)})+\frac{d \Gamma_m (x, h^{(m)})}{d x}\right|_{x=x_*}\delta x
\end{equation}
where $\displaystyle x=(q_*+\Delta)^2 R_G^2=x_*+\delta x$, $\displaystyle\delta x\approx 2 x_* \frac{\Delta}{q_*}$. Comparing (\ref{my_gamma}) with (\ref{real_gamma}) gives
$$
\left.\kappa=\frac{2x_*}{m q_*}\frac{d \Gamma_m (x, h^{(m)})}{d x}\right|_{x=x_*}
$$
Finally,
\begin{equation}
\label{approx_gamma}
\left.\Gamma_m({\mathbf{q}}_1,\dots,{\mathbf{q}}_m)\approx\tilde{\Gamma}_m(\sum_{\alpha=1}^{m}\frac{\Delta_\alpha}{q_*},h^{(m)})=
\Gamma_m(x_*, h^{(m)})+
\frac{2x_*}{m}\frac{d \Gamma_m (x, h^{(m)})}{d x}\right|_{x=x_*} \sum_{\alpha=1}^{m}\frac{\Delta_\alpha}{q_*}
\end{equation}
The angle variables $h^{(m)}$ are calculated from the vectors $q_* {\mathbf{n}}_i$. Here we assumed that the latter sum up to zero, which
is only {\it approximately} true. The approximation (\ref{approx_gamma}) allows us to take into account the small differences in the
length of the wave-vectors from the first and the second shells.

The last step involves the minimization of the free energies with respect to the arbitrary phases $\phi^{(i)}_{j}$. The algebra is
labouring but straightforward. Denoting $a_1$ by $a$, and $a_2$ by $b$ we obtain
\begin{eqnarray}
\label{energies}
F_{FCC}&=&4a^2\Gamma_2(x_*)+3b^2\Gamma_2\left(\frac{4}{3}x_*\right)+12 a^2 b\tilde{\Gamma}_3\left(\frac{2}{\sqrt{3}}-1\right) +\Xi_1^{(F)} a^4+\frac{3}{4}\Xi_2^{(F)} b^4+4\Xi_3^{(F)} a^2 b^2 \nonumber \\
F_{OBDD}&=&6 a^2\Gamma_2(x_*)+4b^2 \Gamma_2\left(\frac{3}{2}x_*\right)+8 a^3 \tilde{\Gamma}_3(0)+\frac{3}{2}\Xi_1^{(O)}a^4+\Xi_2^{(O)}b^4+
2\Xi_3^{(O)}a^2 b^2 \\
F_{GYR}&=&12 a^2 \Gamma_2(x_*)+6 b^2 \Gamma_2\left(\frac{4}{3}x_*\right)+8a^3 \tilde{\Gamma}_3(0)-12 a^2 b \tilde{\Gamma}_3\left(\frac{2}{\sqrt{3}}-1\right) \nonumber \\
&& +8 b^3  \tilde{\Gamma}_3\left(2\sqrt{3}-3\right)+3\Xi_1^{(G)} a^4+\frac{3}{2}\Xi_2^{(G)}b^4+2\Xi_3^{(G)}a^3 b
+2\Xi_4^{(G)}a^2 b^2 \nonumber
\end{eqnarray}
where the coefficients $\Xi$ are given in Appendix C.

The one-harmonic free energies for the LAM, HEX, BCC, SC and SL phases are presented in \cite{Leibler:1980,marques90}. As it was noticed
in \cite{Jones:1994}, we have two choices for the FCC free energy, which arise from the different solutions of the phase minimization
equations. The first solution is presented in (\ref{energies}). The second (NFCC in the  notation of \cite{Jones:1994}) was never found to
be stable in our calculations.

We want to emphasize that the second harmonic approximation was used extensively by many authors \cite{marques90,hamley94,gurovich,Roan:1998,Swift:1992,milner97}. However, almost all of them
used the two-harmonic representation of the density profile for {\it all} phases. On contrary, we apply the criterion 
(\ref{criterion}) and represent only the FCC, OBDD and GYR morphologies in the two-harmonic approximation. This approach was 
also tried in several studies \cite{hamley97,podneks96,Holyst:1999}. The main difference with our work is that we not only keep the complicated angle dependence 
of the vertex functions, but also take into account their dependence on the wave vectors moduli.

\section{Results and discussion}
\setcounter{equation}{0} 

In this section we apply the developed formalism to particular systems. As an example we will treat 
dendrimeric tree-like structures defined as follows. The origin of the molecule is a branching point with $n_1$ A-blocks
of the same length $f_1$ attached to it. All $n_1$ A-blocks are topologically equivalent. They form a so-called first generation
of the molecule. The end of each block is also a branching point with $n_2$ B-blocks of the same length $f_2$ attached to it.
All $n_1 n_2$ B-blocks are topologically equivalent and form the second generation. The third generation again consists
of A-blocks and so on till the $g$-th generation - the highest generation of the molecule. We introduce a $g$-dimensional topology
vector ${\mathbf{n}}=\{n_1,n_2,\dots,n_g\}$. The $i$-th element of this vector defines the number of blocks (A-blocks if $i$ is
odd, and B otherwise) originated from {\it each} block in the $(i-1)$-th generation. The lengths of the blocks within one
generation are necessarily the same but can be different in different generations. It is therefore useful to measure the lengths
of the A-blocks in terms of the length of the first generation $f_A\equiv f_1$ ($f_B\equiv f_2$ for the B-blocks correspondingly). Thus,
we introduce the following $g$-dimensional vector ${\boldsymbol{\tau}}=\{1,1,\tau_3,\dots,\tau_g\}$. The length of the $i$-th generation
is given by
\begin{equation}
f_i=\tau_i\left(f_A \epsilon_i+f_B(1-\epsilon_i) \right)
\end{equation}
where
$$
\epsilon_i=\left\{
\begin{array}{lcl}
1 & &i \text{ is odd} \\
0 & &\text{otherwise}
\end{array}
\right.
$$
If we now introduce the total content of the A-component in the molecule $p$, then the following equations hold:
\begin{eqnarray}
p&=&f_A \sum_{i=1}^{g}\epsilon_i \tau_i \prod_{j=1}^{i}n_j \nonumber \\
1-p&=&f_B \sum_{i=1}^{g}(1-\epsilon_i) \tau_i \prod_{j=1}^{i}n_j
\end{eqnarray} 
and for the given topology (fixed $g$, $\mathbf{n}$ and $\boldsymbol{\tau}$) and the overall composition $p$, $f_A$ and $f_B$ are
determined in a unique way.

\begin{figure}
\begin{picture}(0,300)
\put(-50,0){\epsfig{file=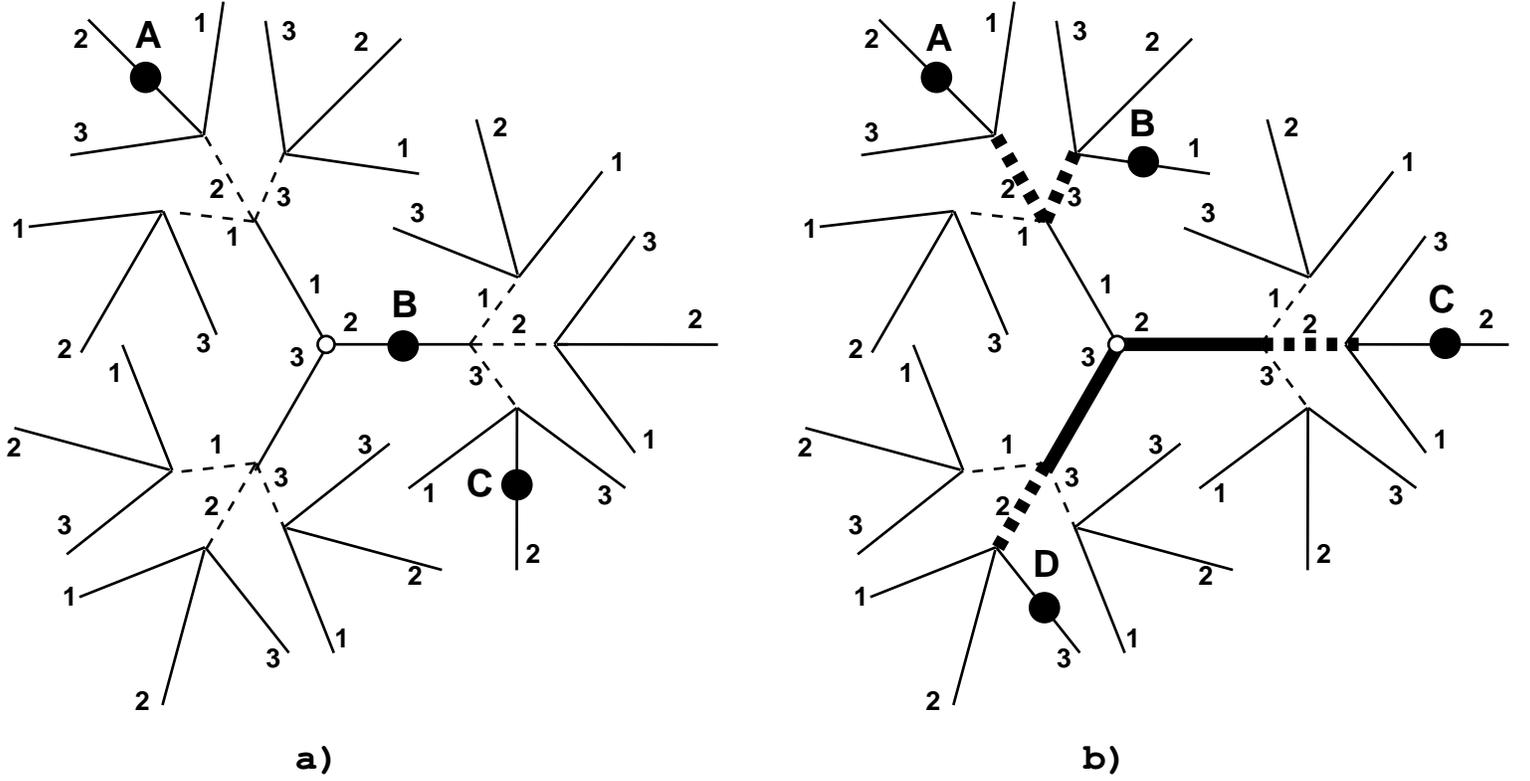,width=20cm}} 
\end{picture}
\caption{Model dendrimeric tree-like structure. The empty circle denotes the origin of the molecule. The filled circles denote the
labeled blocks. a) Here we illustrate the introduced system of coordinates. The positions of the labels A, B and C are
${\mathbf{R}}^{(A)}=\{1,2,2\}$, ${\mathbf{R}}^{(B)}=\{2,0,0\}$ and ${\mathbf{R}}^{(C)}=\{2,3,2\}$, respectively. b) The bold lines
denote the pathways between the labeled blocks. The pathways are: PW(A,B)=\{\{1,2,0\},\{1,3,0\}\} and PW(C,D)=\{\{2,2,0\},\{2,0,0\},\{3,0,0\},\{3,2,0\}\}. } 
\label{dendrimer}
\end{figure}

Next we want to be able to calculate the distance between two given blocks. We introduce the system of coordinates
and number all blocks originated from some block on the $(i-1)$-th generation from $1$ to $n_i$. Then the position of some block is
given by a set of numbers, that define the trajectory from the origin of the molecule to this block, and will be denoted by a
$g$-dimensional vector $\mathbf{R}$. For the $i$-th generation block all components of $\mathbf{R}$ higher than $i$ are equal to zero.
(see Fig.\ref{dendrimer}). The shortest distance between the ends of two blocks with positions ${\mathbf{R}}^{(1)}$ and 
${\mathbf{R}}^{(2)}$ is given by
\begin{equation}
\Delta=\sum_{i=1}^{g}\tau_i \left(f_A \epsilon_i +f_B (1-\epsilon_i)\right)\left(1-\delta_{R^{(1)}_i -R^{(2)}_i,0}\right)\left(2-\delta_{R^{(1)}_i,0}-\delta_{R^{(2)}_i,0}\right)
\end{equation}

We reformulate the framework of Sections 1 \& 2 in the form of
algorithm, depicted schematically as follows:
\begin{enumerate}
\item[1)] Input: $g$, $\mathbf{n}$ and $\boldsymbol{\tau}$
\item[2)] Vary $p$ and $\chi N$ in the ranges of interest and for each point:
\begin{enumerate}
\item[-] calculate the vertex functions
\item[-] calculate the free energies for the mesophases
\item[-] find the stable mesophase (the one with the lowest free energy)
\end{enumerate}
\item[3)] Output: the phase diagram
\end{enumerate}

The calculation of the vertex functions is preformed in the fashion described in Section 1. For example, to calculate
the third order correlation function $G_{AAB}$ we need to move two labels over all A-blocks, and one label over all B-blocks. For 
each configuration of the labels we need to determine the type of diagram and then use the length of the blocks with the labels
for $f_\alpha$, $f_\beta,\dots$ and the distances between those blocks for $\Delta_1$, $\Delta_2,\dots$ in the equations from
Appendix B. The resulting number should be added to $G_{AAB}$.

To distinguish among the different types of diagrams (line, fork, star) we introduce the concept of {\it pathways}. A pathway from
one block to another one consists of the positions of all blocks lying in-between those two (so, the pathway is a set of maximally
$2(g-1)$ vectors). Now suppose we know that the labels (three or four of them) occupy three different blocks. According to 
Appendix A they can form a line- or a star-type diagram. In the first case one of the blocks belongs to the pathway between the 
others, while in the second case it does not. Checking this condition allows us to distinguish among different types of diagrams
(In the 4-block case the number of conditions increases but the general idea remains the same).

The procedure of the free energy minimization with respect to the amplitudes $a$ and $b$ is the same as that of Leibler 
\cite{Leibler:1980} and is not presented here. We proceed with the results for different dendrimers.

\begin{figure}
\begin{picture}(0,350)
\put(-55,0){\epsfig{file=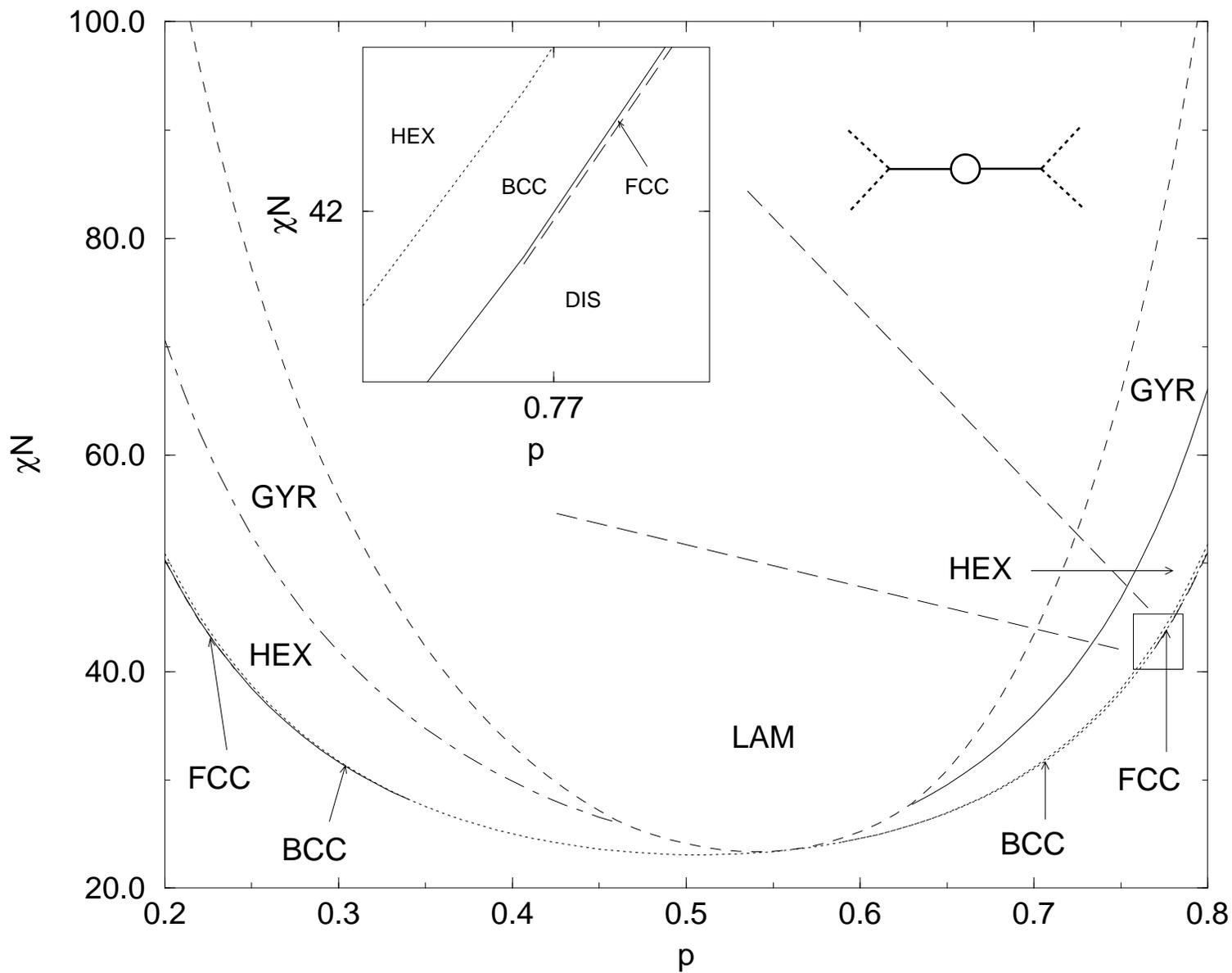,width=20cm}} 
\end{picture}
\caption{Phase diagram for the $g=2$, ${\mathbf{n}}=\{2,2\}$ and $\boldsymbol{\tau}=\{1,1\}$ dendrimer. The DIS-FCC line on the
small graph does not touch the FCC-BCC line because of the finite step size.} 
\label{2_2}
\end{figure}

\begin{figure}
\begin{picture}(0,350)
\put(-55,0){\epsfig{file=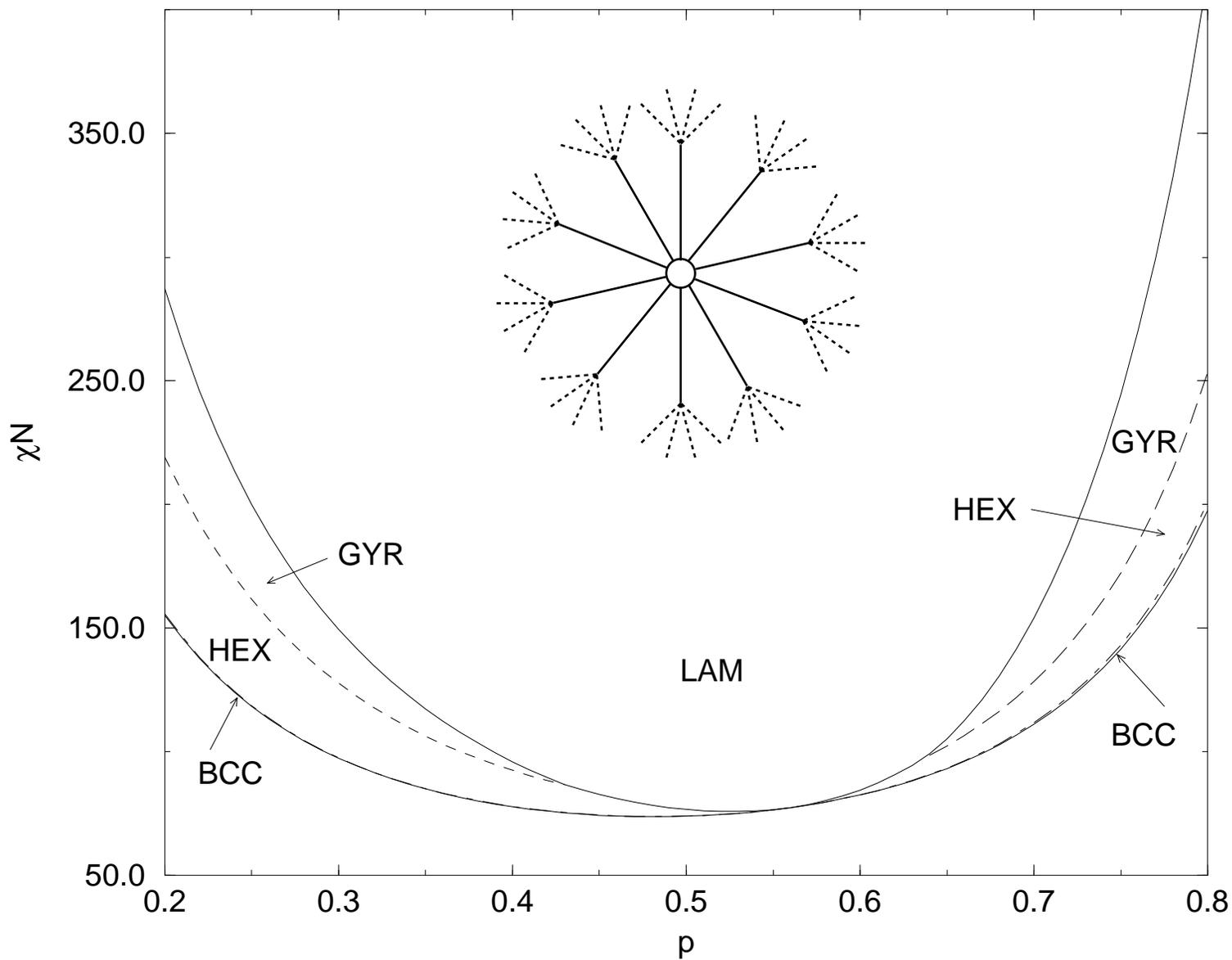,width=20cm}} 
\end{picture}
\caption{Phase diagram for the $g=2$, ${\mathbf{n}}=\{10,4\}$ and $\boldsymbol{\tau}=\{1,1\}$ dendrimer. } 
\label{10_4}
\end{figure}

\begin{figure}
\begin{picture}(0,350)
\put(-55,0){\epsfig{file=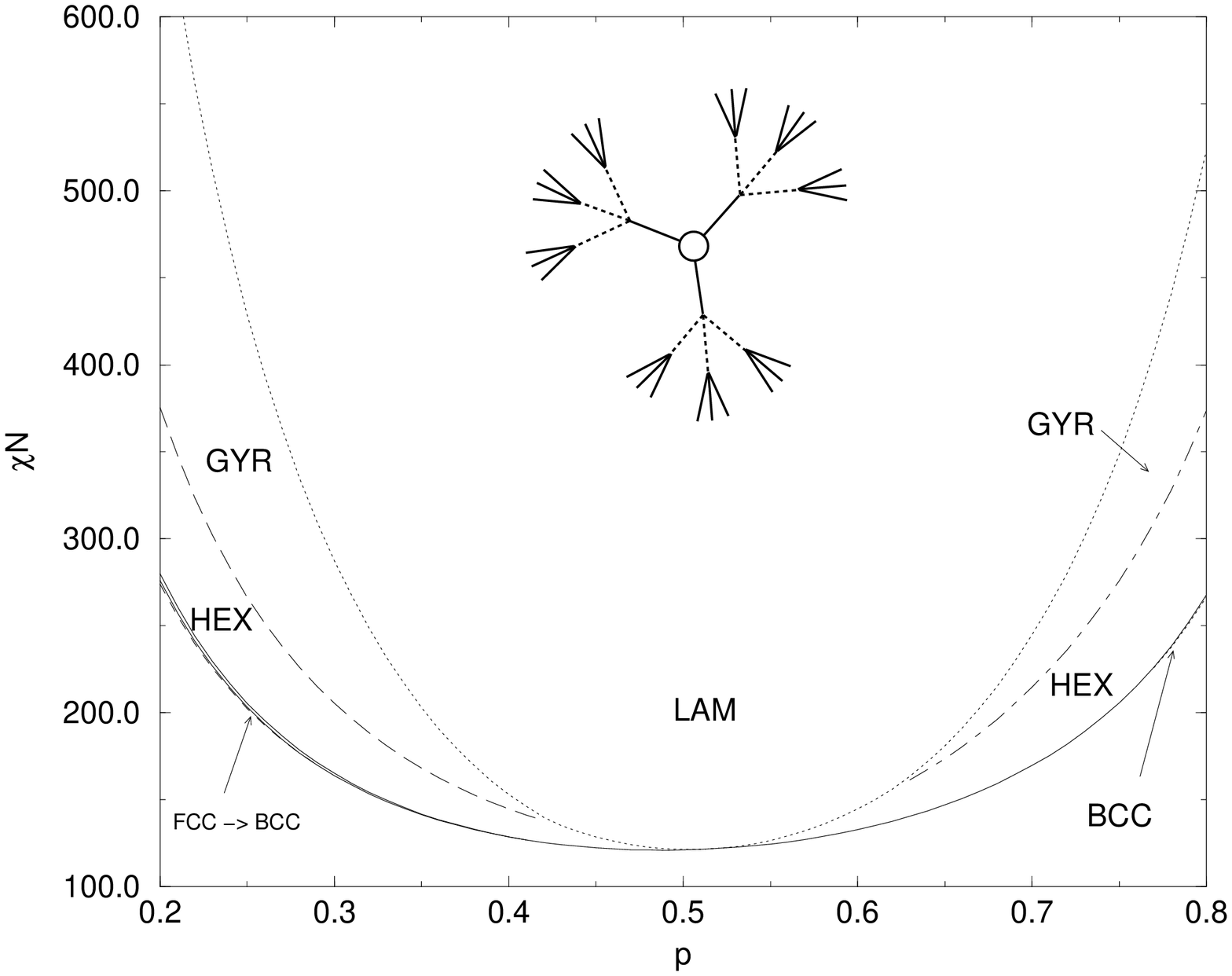,width=20cm}} 
\end{picture}
\caption{Phase diagram for the $g=3$, ${\mathbf{n}}=\{3,3,3\}$ and $\boldsymbol{\tau}=\{1,1,1\}$ dendrimer. } 
\label{3_3_3}
\end{figure}

\begin{figure}
\begin{picture}(0,350)
\put(-55,0){\epsfig{file=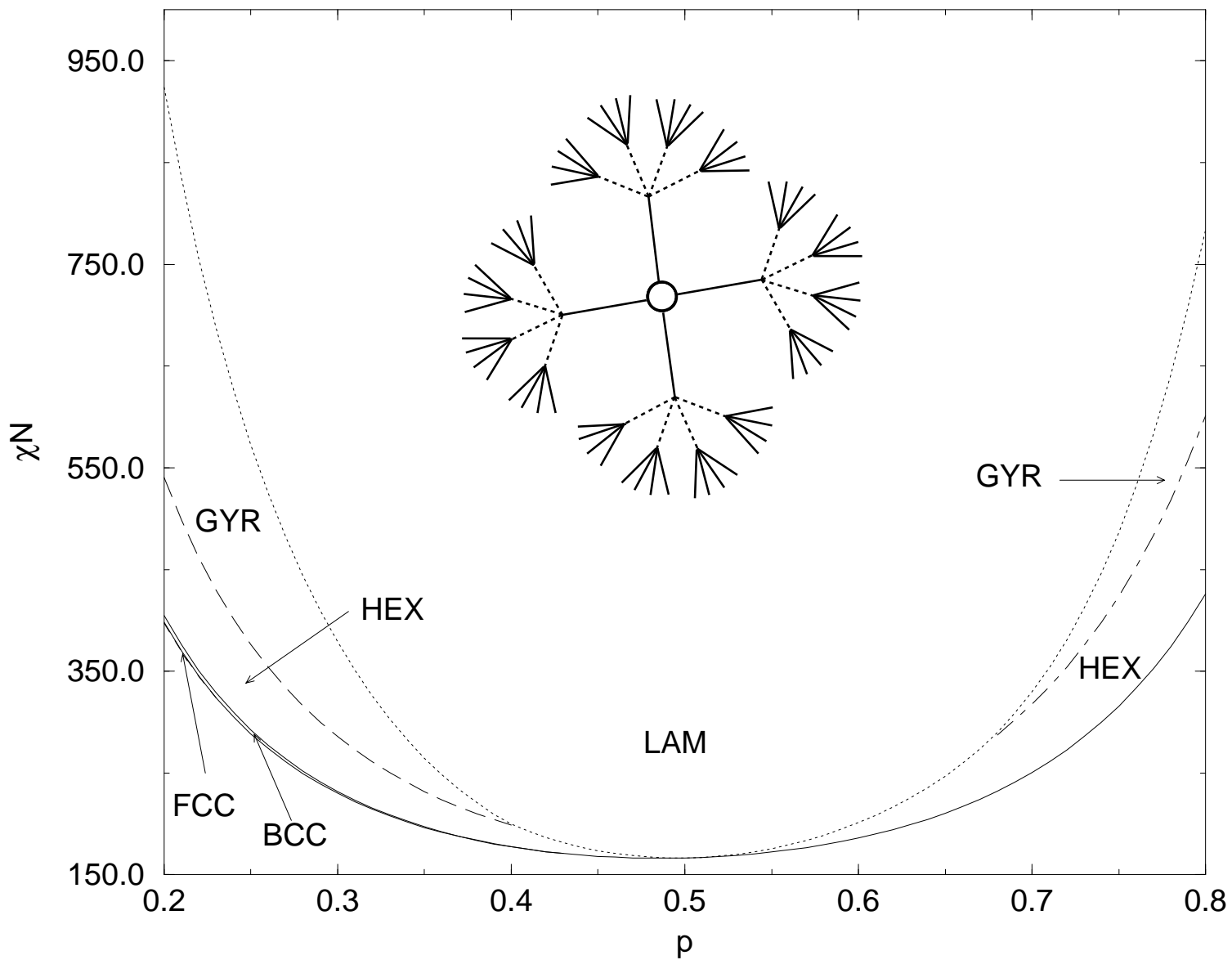,width=20cm}} 
\end{picture}
\caption{Phase diagram for the $g=3$, ${\mathbf{n}}=\{4,4,4\}$ and $\boldsymbol{\tau}=\{1,1,1\}$ dendrimer. The cartoon
of the dendrimer is not in scale: all lengths of the A-blocks should be the same.} 
\label{4_4_4}
\end{figure}

\begin{figure}
\begin{picture}(0,350)
\put(-55,0){\epsfig{file=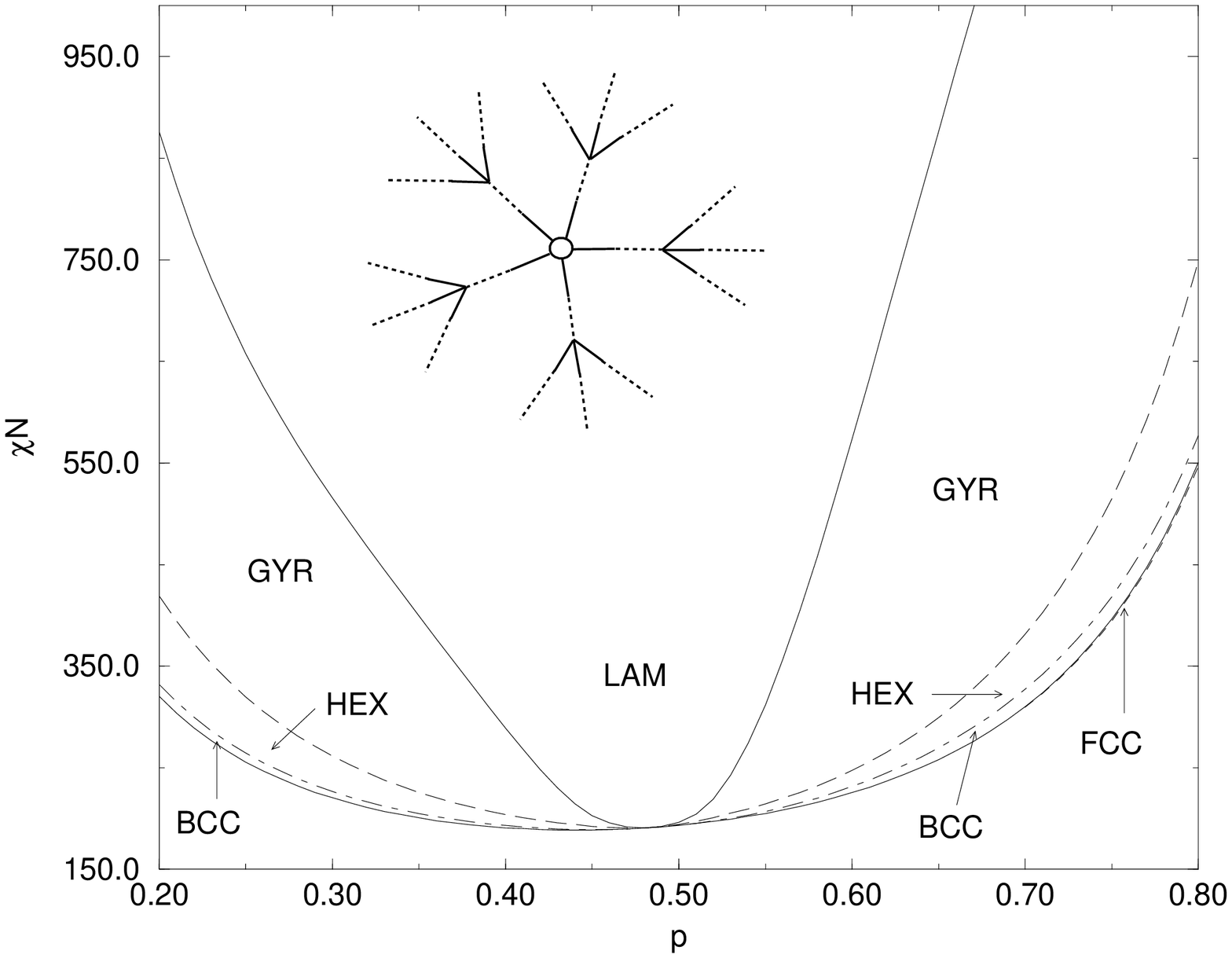,width=20cm}} 
\end{picture}
\caption{Phase diagram for the $g=4$, ${\mathbf{n}}=\{5,1,3,1\}$ and $\boldsymbol{\tau}=\{1,1,1,1\}$ dendrimer.} 
\label{5_1_3_1}
\end{figure}

\begin{figure}
\begin{picture}(0,350)
\put(-55,0){\epsfig{file=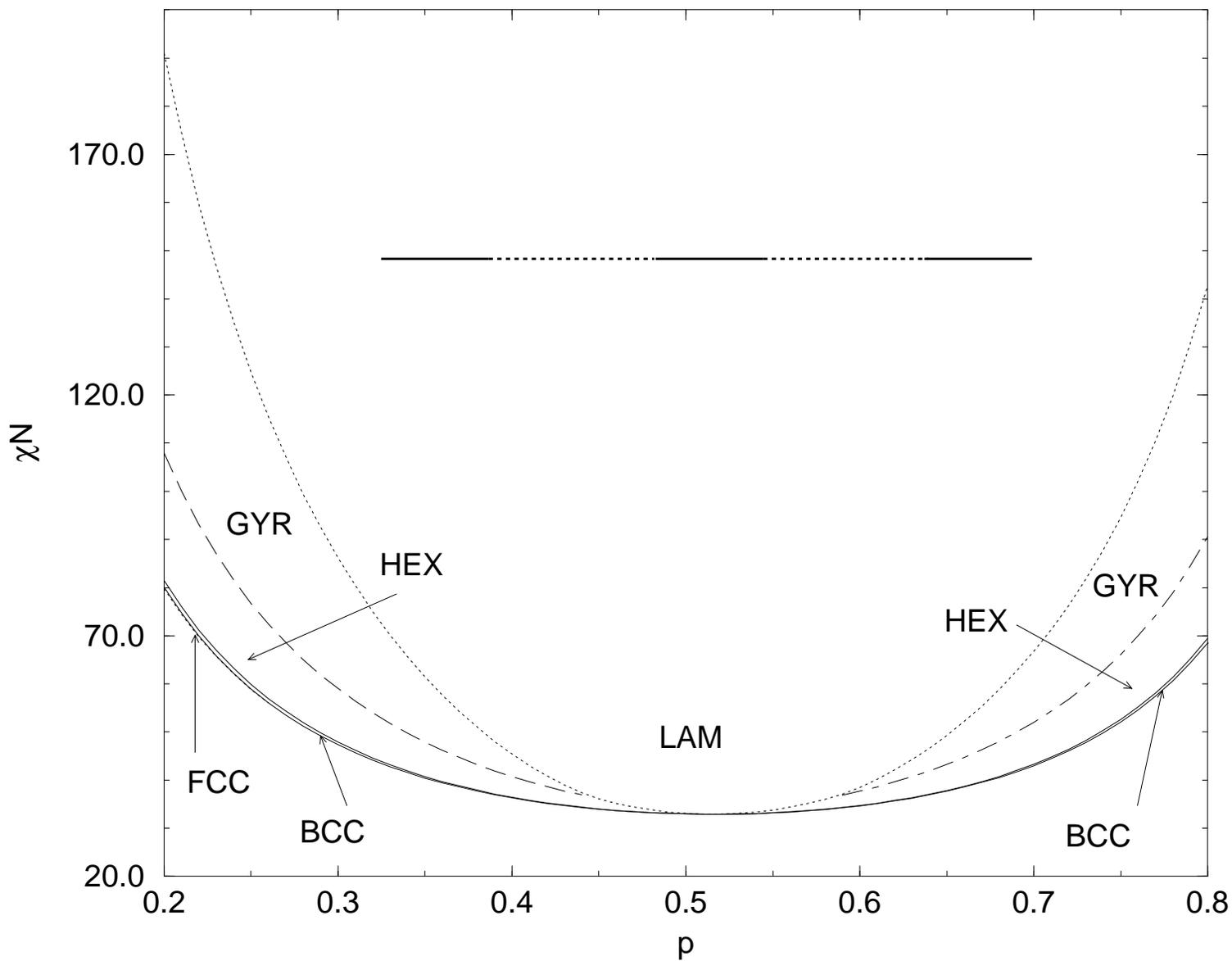,width=20cm}} 
\end{picture}
\caption{Phase diagram for the $g=5$, ${\mathbf{n}}=\{1,1,1,1,1\}$ and $\boldsymbol{\tau}=\{1,1,1,1,1\}$ dendrimer (pentablock).} 
\label{penta}
\end{figure}

Figs.\ref{2_2}-\ref{penta} show the phase behaviour of various architectures. Despite the differences in topology, number of blocks
and branching points, all systems behave in a similar way. Their phase diagrams are very close to those of the diblock 
\cite{Leibler:1980},
triblock \cite{mayes89}, stars and graft copolymers \cite{Cruz:1986,Dobrynin:1993}. The total number of blocks and branching points
influences the spinodal temperature, while the ratio $n_A/n_B$ and the number of branching points influences the asymmetry of phase
diagrams. (A similar observation was made in \cite{Cruz:1986} where the authors connected the increase of complexity of molecules with
the shift of the spinodal temperature.) Moreover, the relative sizes of the stability regions for different phases are almost the same
for all systems, while the types of phases and their order is completely independent of the topology of molecules. Therefore, we
conclude that all mean-field model systems with Gaussian chains will have phase diagrams similar to Figs.\ref{2_2}-\ref{penta}. This 
is in agreement with the recently proposed {\it constituting block copolymer hypothesis} \cite{hadji2000}
which explains the phase behaviour of complicated molecules on the basis of the behaviour of their constitutive blocks. For example, in 
the case of the two-generation dendrimer from Fig.\ref{2_2}, the $\displaystyle\text{A}_1 \text{B}_{2}$ miktoarm-stars are the 
building blocks of the molecules and their behaviour qualitatively corresponds to the phase diagram of the $\{2,2\}$-dendrimer.
Since we can always construct a complicated molecule from simple ones, like diblock, triblock and so on, the resulting phase behaviour  
is going to be typical. However, unlike the general theory of Sections 2 and 3, our model system does not consider structures where
the ends of the blocks are the branching points with both A and B blocks attached to them. (This is necessary, for example, to model
comb copolymers.) Therefore we cannot be sure that such systems will also have typical phase diagrams. There is also evidence of the 
opposite: Dobrynin and Erukhimovich found that the theory fails for the homo-stars $A_nB_n$ with $n>5$ and the corresponding phase 
diagrams contain regions where the phase behaviour cannot be predicted within the used approach \cite{Dobrynin:1993}.
 Nevertheless, the other phase diagrams show the same typical behaviour.

We also want to notice the complete absence of the OBDD morphology. This is not surprising even for the $\{4,4,4\}$-dendrimers 
with the 4-fold symmetry of molecules. As it was shown by Matsen and Bates \cite{matsen97}, the OBDD phase has larger interface in 
comparison with the GYR phase and could be stable only at very low temperatures. It is however screened by other morphologies 
(BCC and HEX).

Another interesting feature is a notably big region of stability of the bicontinuous gyroid phase in Fig.\ref{5_1_3_1}. For such
molecules the highly branched structure and the relative conformational freedom of the end A-B blocks favor the GYR phase to the
LAM phase. This is not a feature of the $\{3,3,3\}$- and $\{4,4,4\}$-dendrimers, for example, because there the outer shell of the molecule consists only of the A blocks. As the AB dendrimers were not studied experimentally yet, we recommend the systems like $\{5,1,3,1\}$
to produce relatively easy the bicontinuous gyroid phase.

At the end we want to mention that our tree-like model is capable of describing stars, polyblocks, umbrellas, and many other architectures. Because of space limitations those results are not presented here.

\section{Conclusion}

In the present paper we have shown how to calculate phase diagrams for systems with arbitrary topology. The developed method is
 a generalization of the work by Leibler \cite{Leibler:1980}. The monomeric correlation functions are calculated with the help of a special
diagrammatic technique which allows us to treat an arbitrary molecular topology. Apart from the assumption that the polymers
obey Gaussian statistics even below the order-disorder transition temperature, the calculation of the vertex functions is
approximation-free. The free energies for different mesophases are calculated within the
second-harmonic approximation as it was formulated by Hamley and Podneks \cite{hamley97,podneks96}. The influence of this approximation and its relationship with the random-phase approximation requires additional study.

The architecture of polymer molecules is known to influence the phase diagram only slightly \cite{Cruz:1986}. This was once
again confirmed by our studies of highly-branched model dendrimers. We found that the topology of molecules influences the spinodal 
temperature and asymmetry of the phase diagrams, but not the types of phases and their order. The typical phase diagrams are presented 
in the Results and Discussion. That model system, however, does not cover all possible architectures. Many more phase diagrams together 
with the C-code used to obtain them can be found on the Internet: {\it http://rugmd4.chem.rug.nl/\~{}morozov/research.html}.

Our method can be a starting point for a fluctuation theory of systems with arbitrary topology. This will require some
straightforward generalization of the Hartree approximation by Brazovskii \cite{Brazovskii:1975,Brazovskii:1987} and Fredrickson and Helfand \cite{Fredrickson:1987}, since for
some systems (see, for example, Figs.4 and 5) the 4-th order vertex function $\Gamma_4(h_1,h_2)$ exhibits a strong angle-dependence
and, therefore, cannot be approximated by the constant value $\Gamma_4(0,0)$ or any other. One of the possible ways for such a generalization
was proposed by Dobrynin and Erukhimovich \cite{Dobrynin:1991}.

\section{Acknowledgments}

We thank A. Zvelindovsky, M. Reenders and I. Reviakine for reading the manuscript.

\newpage

\renewcommand{\theequation}{\thesection.\arabic{equation}}

\appendix
\section{Diagrammatic technique}

We consider a polymer molecule consisting of $n_A$ blocks of type A and $n_B$ blocks of type B. The total
number of monomers in the molecule is $N$. We shall use Greek letters to number the blocks: 
$\alpha,\beta,\gamma,\delta=1\dots n_A+n_B$. The length of the $\alpha$-th block is $L_\alpha=f_\alpha N$. We
use solid lines to denote blocks and dashed lines to denote arbitrary distance between blocks. The circles 
denote the labeled monomers. We use $s,s',s''\dots$ for the positions of the labels on their blocks. 
A diagram  with $n$ labels will be called the $n$-th order diagram. The labels act as {\it sources} of momentum. 
A circle labeled with $a,b,c\dots$ emits ${\mathbf{q}}_a,{\mathbf{q}}_b,{\mathbf{q}}_c \dots$ respectively.
The total momentum of any
diagram should be zero, therefore all external lines should carry zero momentum. 
We also introduce a convention (which does not affect our results) that all momenta flow from left to right 
along the chain. This can be illustrated by the following:

\begin{picture}(100,80)
\linethickness{1.pt}
\put(30,40){\line(1,0){200}}
\put(60,40){\circle*{4}}
\put(130,40){\circle*{4}}
\put(200,40){\circle*{4}}
\linethickness{.5pt}
\drawline[0](35,50)(55,50)
\drawline[2](52,51)(55,50)(52,49)
\put(40,55){\makebox(10,10)[c]{${\mathbf{0}}$}}
\drawline[0](75,50)(115,50)
\drawline[2](112,51)(115,50)(112,49)
\drawline[0](145,50)(185,50)
\drawline[2](182,51)(185,50)(182,49)
\drawline[0](205,50)(225,50)
\drawline[2](222,51)(225,50)(222,49)
\put(90,55){\makebox(10,10)[c]{${\mathbf{q}}_1$}}
\put(160,55){\makebox(10,10)[c]{${\mathbf{q}}_1+{\mathbf{q}}_2$}}
\put(210,55){\makebox(10,10)[c]{${\mathbf{0}}$}}
\drawline[0](60,10)(60,30)
\drawline[0](130,10)(130,30)
\drawline[0](200,10)(200,30)
\drawline[2](59,27)(60,30)(61,27)
\drawline[2](129,27)(130,30)(131,27)
\drawline[2](199,27)(200,30)(201,27)
\put(47,15){\makebox(10,10)[c]{${\mathbf{q}}_1$}}
\put(117,15){\makebox(10,10)[c]{${\mathbf{q}}_2$}}
\put(187,15){\makebox(10,10)[c]{${\mathbf{q}}_3$}}
\end{picture}

A given number of blocks can form topologically different diagrams:

\begin{picture}(100,60)
\linethickness{.5pt}

\put(0,40){\makebox(2,2)[c]{2:}}
\drawline[0](20,40)(30,40)
\dashline{4}(31,40)(40,40)
\drawline[0](41,40)(50,40)

\put(100,40){\makebox(2,2)[c]{3:}}
\drawline[0](120,40)(130,40)
\dashline{4}(131,40)(140,40)
\drawline[0](141,40)(150,40)
\dashline{4}(151,40)(160,40)
\drawline[0](161,40)(170,40)

\drawline[0](200,40)(210,40)
\dashline{4}(211,40)(230,40)
\drawline[0](231,40)(240,40)
\dashline{4}(220.5,40)(220.5,50)
\drawline[0](220.5,51)(220.5,60)

\put(35,44){\makebox(2,2)[c]{$\scriptstyle line$}}
\put(145,44){\makebox(2,2)[c]{$\scriptstyle line$}}
\put(208,43){\makebox(2,2)[c]{$\scriptstyle star$}}

\put(0,15){\makebox(2,2)[c]{4:}}
\drawline[0](30,15)(40,15)
\dashline{4}(41,15)(50,15)
\drawline[0](51,15)(60,15)
\dashline{4}(61,15)(70,15)
\drawline[0](71,15)(80,15)
\dashline{4}(81,15)(90,15)
\drawline[0](91,15)(100,15)

\drawline[0](130,15)(140,15)
\dashline{4}(141,15)(150,15)
\drawline[0](151,15)(160,15)
\dashline{4}(161,15)(170,15)
\dashline{4}(170,10)(170,20)
\drawline[0](170,0)(170,9)
\drawline[0](170,21)(170,30)

\dashline{4}(200,15)(240,15)
\dashline{4}(200,10)(200,20)
\dashline{4}(240,10)(240,20)
\drawline[0](200,0)(200,9)
\drawline[0](200,21)(200,30)
\drawline[0](240,0)(240,9)
\drawline[0](240,21)(240,30)

\put(65,19){\makebox(2,2)[c]{$\scriptstyle line$}}
\put(151,18){\makebox(2,2)[c]{$\scriptstyle fork$}}
\put(219,18){\makebox(2,2)[c]{$\scriptstyle star$}}

\end{picture}

Necessarily, all momenta flowing in and out of each branching point should add up to zero.

Following \cite{Leibler:1980} we assume that all wavevectors have the same length $|{\mathbf{q}}|=q_*$
and introduce the following angle variables:
\begin{eqnarray}
&&\text{for the 3rd order diagrams:}\quad {\mathbf{q}}_i^2=q_*^2 h^{(3)}_{i}\quad , \quad h^{(3)}=(1,1,h) \nonumber \\  
&&\text{for the 4th order diagrams:}\quad ({\mathbf{q}}_i + {\mathbf{q}}_j)^2=q_*^2 h^{(4)}_{ij} \nonumber 
\end{eqnarray}
where
$$
h^{(4)}=\left(
\begin{array}{cccc}
4 & h_1 & 4-h_1-h_2 & h_2\\
h_1 & 4 & h_2 & 4-h_1-h_2\\
4-h_1-h_2 & h_2 & 4 & h_1\\
h_2 & 4-h_1-h_2 & h_1 & 4
\end{array}
\right)
$$

The matrix $h^{(4)}$ has two important symmetry properties: $h^{(4)}_{ij}=h^{(4)}_{ji}$ and
$h^{(4)}_{ij}=h^{(4)}_{kl}$ for $i\neq j \neq k \neq l$. These properties will be extensively
used in Appendix B.

For any internal line we subscribe the Gaussian propagator:
$$
{\mathcal{P}}(l)=e^{-\frac{x}{N}l}\quad\text{,}\qquad x=q^2_* R_G^2
$$
where $R_G$ is the radius of gyration of an ideal chain of $N$ monomers. The product of the propagators
is integrated over all possible positions of all labels on their blocks. The result should be then
multiplied with the symmetry factor arising from all possible permutations of labels that result in
topologically the same diagram and keep labels on the blocks they initially occupied. Finally, the 
normalization constant $1/N$ is added.

The following functions will appear as the building blocks for all diagrams:
\begin{eqnarray}
F_f&=&\frac{1-e^{-f x}}{x} \nonumber \\
{\mathcal{F}}_f(h)&=&\left[\quad\frac{F_{f h}}{h}\quad|\quad F_f\quad|\quad f \quad\right] \nonumber \\
J_f (h)&=&\frac{1}{x^2}\left[\quad \frac{1}{h}-\frac{e^{-f x}}{h-1}-\frac{e^{-h f x}}{h-h^2}\quad|\quad
1-e^{-f x}-f x e^{-f x} \quad|\quad f x + e^{-f x} -1\quad\right] \nonumber \\
{\mathcal{D}}_{f}(h)&=&\frac{1}{x^2}\left[\quad \frac{h f x+e^{-h f x}-1}{h^2}\quad|\quad f x+e^{-f x}-1 \quad|\quad \frac{f^2 x^2}{2}\quad \right] \nonumber \\
T_1(f,h)&=&\frac{1}{x^4}\left[\quad\frac{fx}{h}+e^{-fx}\frac{fx}{h-1}+e^{-fx}\frac{2h-3}{(h-1)^2} +\frac{e^{-fhx}}{h^2(h-1)^2}-\frac{2h+1}{h^2} \quad|\quad \right. \nonumber \\ 
&& \left. \frac{f^2 x^2}{2} e^{-fx}+2 fx e^{-fx}+fx-3+3e^{-fx} \quad|\quad  \frac{f^2 x^2}{2}+3-2fx-(3+fx)e^{-fx}\quad\right] \nonumber
\end{eqnarray}
\begin{eqnarray}
T_2(f,h)&=&\frac{1}{x^3}\left[\quad\frac{1}{h}-\frac{e^{-fhx}}{h(h-1)^2}-\frac{h-2}{(h-1)^2}e^{-fx}-\frac{fx}{h-1}e^{-fx}\quad|\quad \right.\nonumber \\
&&\qquad \left. 1-e^{-fx}-f x e^{-fx}- \frac{f^2 x^2}{2}e^{-fx}\quad|\quad 2e^{-fx}+f x e^{-fx}+f x-2 \quad\right] \nonumber
\end{eqnarray}
All functions of the angle variable $h$ are defined in the following way:
$$
\dots(h)=\left[\quad h\neq0,1 \quad|\quad h=1\quad|\quad h=0 \quad\right]
$$
Some of these functions were previously defined in \cite{Leibler:1980,mayes89,mayes:phd}. Separate elements of the presented
diagrammatic technique were used in \cite{Roan:1998,henk,slot,rik}.

\section{Calculation of diagrams}

\begin{picture}(100,50)
\put(0,20){\makebox(10,10)[c]{$\displaystyle S_{ij}=$}}
\linethickness{1.pt}
\drawline[0](20,25)(50,25)
\put(30,25){\circle*{3.2}}
\put(40,25){\circle*{3.2}}
\put(21,21){\makebox(2,2)[c]{$\scriptscriptstyle \alpha$}}
\put(39,29){\makebox(2,2)[c]{$\scriptstyle s$}}
\put(30,30){\makebox(2,2)[c]{$\scriptstyle s'$}}
\put(55,20){\makebox(10,10)[c]{$\displaystyle +$}}

\drawline[0](70,25)(90,25)
\dashline{4}(91,25)(110,25)
\drawline[0](111,25)(130,25)
\put(80,25){\circle*{3.2}}
\put(120,25){\circle*{3.2}}
\put(71,21){\makebox(2,2)[c]{$\scriptscriptstyle \beta$}}
\put(111,21){\makebox(2,2)[c]{$\scriptscriptstyle \alpha$}}
\put(119,29){\makebox(2,2)[c]{$\scriptstyle s$}}
\put(80,30){\makebox(2,2)[c]{$\scriptstyle s'$}}
\put(99,29){\makebox(2,2)[c]{$\scriptstyle N\Delta$}}
\end{picture}

\begin{eqnarray}
I)&=&\frac{2}{N}\int_0^{L_\alpha}ds\int_0^{s}ds'\,{\mathcal{P}}\left(s-s'\right)=2 N {\mathcal{D}}_{f_\alpha}(1) \nonumber \\
II)&=&\frac{1}{N}\int_0^{L_\alpha}ds\int_0^{L_\beta}ds'\,{\mathcal{P}}\left(s+s'+N\Delta\right)=N F_{f_\alpha}F_{f_\beta}e^{-x \Delta}\nonumber 
\end{eqnarray}

\begin{picture}(100,100)
\put(0,75){\makebox(10,10)[c]{$\displaystyle G_{ijk}=$}}
\linethickness{1.pt}
\drawline[0](20,80)(50,80)
\put(27,80){\circle*{3.2}}
\put(35,80){\circle*{3.2}}
\put(43,80){\circle*{3.2}}
\put(21,76){\makebox(2,2)[c]{$\scriptscriptstyle \alpha$}}
\put(42,84){\makebox(2,2)[c]{$\scriptstyle s$}}
\put(35,85){\makebox(2,2)[c]{$\scriptstyle s'$}}
\put(28,85){\makebox(2,2)[c]{$\scriptstyle s''$}}

\put(55,75){\makebox(10,10)[c]{$\displaystyle +$}}

\drawline[0](70,80)(90,80)
\dashline{4}(91,80)(110,80)
\drawline[0](111,80)(130,80)
\put(75,80){\circle*{3.2}}
\put(85,80){\circle*{3.2}}
\put(120,80){\circle*{3.2}}
\put(70,76){\makebox(2,2)[c]{$\scriptscriptstyle \beta$}}
\put(112,76){\makebox(2,2)[c]{$\scriptscriptstyle \alpha$}}
\put(119,84){\makebox(2,2)[c]{$\scriptstyle s$}}
\put(85,85){\makebox(2,2)[c]{$\scriptstyle s'$}}
\put(76,85){\makebox(2,2)[c]{$\scriptstyle s''$}}
\put(99,84){\makebox(2,2)[c]{$\scriptstyle N\Delta$}}
\put(119,74){\makebox(2,2)[c]{$\scriptstyle a$}}

\put(135,75){\makebox(10,10)[c]{$\displaystyle +$}}

\drawline[0](150,80)(170,80)
\dashline{4}(171,80)(190,80)
\drawline[0](191,80)(210,80)
\dashline{4}(211,80)(230,80)
\drawline[0](231,80)(250,80)
\put(160,80){\circle*{3.2}}
\put(200,80){\circle*{3.2}}
\put(240,80){\circle*{3.2}}
\put(151,76){\makebox(2,2)[c]{$\scriptscriptstyle \gamma$}}
\put(191,76){\makebox(2,2)[c]{$\scriptscriptstyle \beta$}}
\put(232,76){\makebox(2,2)[c]{$\scriptscriptstyle \alpha$}}
\put(239,84){\makebox(2,2)[c]{$\scriptstyle s$}}
\put(200,85){\makebox(2,2)[c]{$\scriptstyle s'$}}
\put(161,85){\makebox(2,2)[c]{$\scriptstyle s''$}}
\put(179,84){\makebox(2,2)[c]{$\scriptstyle N\Delta_1$}}
\put(219,84){\makebox(2,2)[c]{$\scriptstyle N\Delta_2$}}
\put(239,74){\makebox(2,2)[c]{$\scriptstyle a$}}
\put(199,74){\makebox(2,2)[c]{$\scriptstyle b$}}
\put(159,74){\makebox(2,2)[c]{$\scriptstyle c$}}

\put(80,15){\makebox(10,10)[c]{$\displaystyle +$}}

\drawline[0](95,20)(115,20)
\dashline{4}(116,20)(155,20)
\drawline[0](156,20)(175,20)
\put(105,20){\circle*{3.2}}
\put(165,20){\circle*{3.2}}
\dashline{4}(135.5,20)(135.5,40)
\drawline[0](135.5,41)(135.5,60)
\put(135.5,50){\circle*{3.2}}
\put(96,16){\makebox(2,2)[c]{$\scriptscriptstyle \alpha$}}
\put(157,16){\makebox(2,2)[c]{$\scriptscriptstyle \beta$}}
\put(137,59){\makebox(2,2)[c]{$\scriptscriptstyle \gamma$}}

\put(104,24){\makebox(2,2)[c]{$\scriptstyle s$}}
\put(166,25){\makebox(2,2)[c]{$\scriptstyle s'$}}
\put(141,50){\makebox(2,2)[c]{$\scriptstyle s''$}}

\put(124,23){\makebox(2,2)[c]{$\scriptstyle N\Delta_1$}}
\put(145,23){\makebox(2,2)[c]{$\scriptstyle N\Delta_2$}}
\put(143,35){\makebox(2,2)[c]{$\scriptstyle N\Delta_3$}}

\put(104,14){\makebox(2,2)[c]{$\scriptstyle a$}}
\put(164,14){\makebox(2,2)[c]{$\scriptstyle b$}}
\put(129,49){\makebox(2,2)[c]{$\scriptstyle c$}}

\end{picture}

\begin{eqnarray}
I)&=&\frac{1}{N}\sum_{a\neq b \neq c =1}^{3}\int_{0}^{L_\alpha}ds \int_{0}^{s}ds' \int_{0}^{s'}ds''\,{\mathcal{P}}\left(h^{(3)}_a(s-s')+h^{(3)}_c(s'-s'')\right)\nonumber \\
&&\qquad \qquad =2N^2 \left[2g_2(f_\alpha,h)+g_2(f_\alpha,1)\right]\nonumber \\
II)&=&\frac{1}{N}\int_{0}^{L_\alpha}ds \int_{0}^{L_\beta}ds' \int_{0}^{s'}ds'' \left[ {\mathcal{P}}\left(h^{(3)}_c(s'-s'')+h^{(3)}_a (L_\beta -s'+N \Delta +s)\right) \right. \nonumber \\
&&\qquad\qquad+\left.{\mathcal{P}}\left(h^{(3)}_b(s'-s'')+h^{(3)}_a (L_\beta -s'+N \Delta +s)\right) \right]\nonumber \\
&&=N^2 e^{-h^{(3)}_a x \Delta} {\mathcal{F}}_{f_\alpha}(h^{(3)}_a)\left[J_{f_\beta}(h)+J_{f_\beta}(h^{(3)}_a) \right]\nonumber
\end{eqnarray}
\begin{eqnarray}
III)&=&\frac{1}{N}\int_{0}^{L_\alpha}ds \int_{0}^{L_\beta}ds' \int_{0}^{L_\gamma}ds''\,{\mathcal{P}}\left(h^{(3)}_c (s''+N\Delta_1+s') +h^{(3)}_a(L_\beta-s'+N\Delta_2+s)\right) \nonumber \\
&&=N^2 e^{-x f_\beta h^{(3)}_c} e^{-x(h^{(3)}_c \Delta_1+h^{(3)}_a \Delta_2)} {\mathcal{F}}_{f_\alpha}(h^{(3)}_a) {\mathcal{F}}_{f_\beta}(h^{(3)}_a - h^{(3)}_c){\mathcal{F}}_{f_\gamma}(h^{(3)}_c) \nonumber \\
IV)&=&\frac{1}{N}\int_{0}^{L_\alpha}ds \int_{0}^{L_\beta}ds' \int_{0}^{L_\gamma}ds''\,{\mathcal{P}}\left(h^{(3)}_a(s+N\Delta_1)+h^{(3)}_b(s'+N\Delta_2)+h^{(3)}_c(s''+N\Delta_3) \right)\nonumber \\
&&=N^2 {\mathcal{F}}_{f_\alpha}(h^{(3)}_a){\mathcal{F}}_{f_\beta}(h^{(3)}_b){\mathcal{F}}_{f_\gamma}(h^{(3)}_c) e^{-x(h^{(3)}_a \Delta_1+h^{(3)}_b \Delta_2+h^{(3)}_c \Delta_3)}\nonumber
\end{eqnarray}

\begin{picture}(100,100)
\put(0,75){\makebox(10,10)[c]{$\displaystyle G_{ijkl}=$}}
\linethickness{1.pt}
\drawline[0](20,80)(50,80)
\put(26,80){\circle*{3.2}}
\put(32,80){\circle*{3.2}}
\put(38,80){\circle*{3.2}}
\put(44,80){\circle*{3.2}}
\put(20,76){\makebox(2,2)[c]{$\scriptscriptstyle \alpha$}}
\put(44,84){\makebox(2,2)[c]{$\scriptstyle s$}}
\put(39,85){\makebox(2,2)[c]{$\scriptstyle s'$}}
\put(33,85){\makebox(2,2)[c]{$\scriptstyle s''$}}
\put(26,85){\makebox(2,2)[c]{$\scriptstyle s'''$}}

\put(55,75){\makebox(10,10)[c]{$\displaystyle +$}}
\drawline[0](70,80)(90,80)
\dashline{4}(91,80)(110,80)
\drawline[0](111,80)(130,80)
\put(75,80){\circle*{3.2}}
\put(80,80){\circle*{3.2}}
\put(85,80){\circle*{3.2}}
\put(120,80){\circle*{3.2}}
\put(70,76){\makebox(2,2)[c]{$\scriptscriptstyle \beta$}}
\put(112,76){\makebox(2,2)[c]{$\scriptscriptstyle \alpha$}}
\put(119,84){\makebox(2,2)[c]{$\scriptstyle s$}}
\put(86,85){\makebox(2,2)[c]{$\scriptstyle s'$}}
\put(81,85){\makebox(2,2)[c]{$\scriptstyle s''$}}
\put(75,85){\makebox(2,2)[c]{$\scriptstyle s'''$}}
\put(99,84){\makebox(2,2)[c]{$\scriptstyle N\Delta$}}
\put(119,74){\makebox(2,2)[c]{$\scriptstyle a$}}

\put(135,75){\makebox(10,10)[c]{$\displaystyle +$}}

\drawline[0](150,80)(170,80)
\dashline{4}(171,80)(190,80)
\drawline[0](191,80)(210,80)
\put(155,80){\circle*{3.2}}
\put(165,80){\circle*{3.2}}
\put(195,80){\circle*{3.2}}
\put(205,80){\circle*{3.2}}
\put(150,76){\makebox(2,2)[c]{$\scriptscriptstyle \beta$}}
\put(191,76){\makebox(2,2)[c]{$\scriptscriptstyle \alpha$}}
\put(204,84){\makebox(2,2)[c]{$\scriptstyle s$}}
\put(195,85){\makebox(2,2)[c]{$\scriptstyle s'$}}
\put(166,85){\makebox(2,2)[c]{$\scriptstyle s''$}}
\put(155,85){\makebox(2,2)[c]{$\scriptstyle s'''$}}
\put(179,84){\makebox(2,2)[c]{$\scriptstyle N\Delta$}}
\put(199,74){\makebox(2,2)[c]{$\scriptstyle b \leftrightarrow a$}}

\put(215,75){\makebox(10,10)[c]{$\displaystyle +$}}

\drawline[0](230,80)(250,80)
\dashline{4}(251,80)(270,80)
\drawline[0](271,80)(290,80)
\dashline{4}(291,80)(310,80)
\drawline[0](311,80)(330,80)
\put(240,80){\circle*{3.2}}
\put(280,80){\circle*{3.2}}
\put(316,80){\circle*{3.2}}
\put(324,80){\circle*{3.2}}
\put(311,76){\makebox(2,2)[c]{$\scriptscriptstyle \alpha$}}
\put(271,76){\makebox(2,2)[c]{$\scriptscriptstyle \beta$}}
\put(231,76){\makebox(2,2)[c]{$\scriptscriptstyle \gamma$}}
\put(323,84){\makebox(2,2)[c]{$\scriptstyle s$}}
\put(316,85){\makebox(2,2)[c]{$\scriptstyle s'$}}
\put(281,85){\makebox(2,2)[c]{$\scriptstyle s''$}}
\put(241,85){\makebox(2,2)[c]{$\scriptstyle s'''$}}
\put(259,84){\makebox(2,2)[c]{$\scriptstyle N\Delta_1$}}
\put(299,84){\makebox(2,2)[c]{$\scriptstyle N\Delta_2$}}
\put(279,74){\makebox(2,2)[c]{$\scriptstyle a$}}
\put(239,74){\makebox(2,2)[c]{$\scriptstyle b$}}

\put(35,25){\makebox(10,10)[c]{$\displaystyle +$}}

\drawline[0](50,30)(70,30)
\dashline{4}(71,30)(90,30)
\drawline[0](91,30)(110,30)
\dashline{4}(111,30)(130,30)
\drawline[0](131,30)(150,30)
\put(60,30){\circle*{3.2}}
\put(96,30){\circle*{3.2}}
\put(104,30){\circle*{3.2}}
\put(140,30){\circle*{3.2}}
\put(132,26){\makebox(2,2)[c]{$\scriptscriptstyle \alpha$}}
\put(91,26){\makebox(2,2)[c]{$\scriptscriptstyle \beta$}}
\put(51,26){\makebox(2,2)[c]{$\scriptscriptstyle \gamma$}}
\put(139,34){\makebox(2,2)[c]{$\scriptstyle s$}}
\put(104,35){\makebox(2,2)[c]{$\scriptstyle s'$}}
\put(97,35){\makebox(2,2)[c]{$\scriptstyle s''$}}
\put(61,35){\makebox(2,2)[c]{$\scriptstyle s'''$}}
\put(79,34){\makebox(2,2)[c]{$\scriptstyle N\Delta_1$}}
\put(119,34){\makebox(2,2)[c]{$\scriptstyle N\Delta_2$}}
\put(139,24){\makebox(2,2)[c]{$\scriptstyle a$}}
\put(59,24){\makebox(2,2)[c]{$\scriptstyle b$}}

\put(155,25){\makebox(10,10)[c]{$\displaystyle +$}}
\drawline[0](170,30)(190,30)
\dashline{4}(191,30)(230,30)
\drawline[0](231,30)(250,30)
\dashline{4}(210.5,30)(210.5,50)
\drawline[0](210.5,51)(210.5,70)
\put(180,30){\circle*{3.2}}
\put(240,30){\circle*{3.2}}
\put(210.5,56){\circle*{3.2}}
\put(210.5,64){\circle*{3.2}}
\put(171,26){\makebox(2,2)[c]{$\scriptscriptstyle \alpha$}}
\put(232,26){\makebox(2,2)[c]{$\scriptscriptstyle \beta$}}
\put(212,69){\makebox(2,2)[c]{$\scriptscriptstyle \gamma$}}
\put(179,34){\makebox(2,2)[c]{$\scriptstyle s$}}
\put(240,35){\makebox(2,2)[c]{$\scriptstyle s'$}}
\put(216,65){\makebox(2,2)[c]{$\scriptstyle s''$}}
\put(216,56){\makebox(2,2)[c]{$\scriptstyle s'''$}}
\put(199,24){\makebox(2,2)[c]{$\scriptstyle N\Delta_1$}}
\put(220,24){\makebox(2,2)[c]{$\scriptstyle N\Delta_2$}}
\put(218,41){\makebox(2,2)[c]{$\scriptstyle N\Delta_3$}}
\put(179,24){\makebox(2,2)[c]{$\scriptstyle a$}}
\put(239,24){\makebox(2,2)[c]{$\scriptstyle b$}}

\end{picture}

\begin{picture}(100,150)
\linethickness{1.pt}
\put(65,145){\makebox(10,10)[c]{$\displaystyle +$}}
\drawline[0](80,150)(100,150)
\dashline{4}(101,150)(120,150)
\drawline[0](121,150)(140,150)
\dashline{4}(141,150)(160,150)
\drawline[0](161,150)(180,150)
\dashline{4}(181,150)(200,150)
\drawline[0](201,150)(220,150)
\put(90,150){\circle*{3.2}}
\put(130,150){\circle*{3.2}}
\put(170,150){\circle*{3.2}}
\put(210,150){\circle*{3.2}}
\put(202,146){\makebox(2,2)[c]{$\scriptscriptstyle \alpha$}}
\put(161,146){\makebox(2,2)[c]{$\scriptscriptstyle \beta$}}
\put(122,146){\makebox(2,2)[c]{$\scriptscriptstyle \gamma$}}
\put(81,146){\makebox(2,2)[c]{$\scriptscriptstyle \delta$}}
\put(209,154){\makebox(2,2)[c]{$\scriptstyle s$}}
\put(170,155){\makebox(2,2)[c]{$\scriptstyle s'$}}
\put(131,155){\makebox(2,2)[c]{$\scriptstyle s''$}}
\put(91,155){\makebox(2,2)[c]{$\scriptstyle s'''$}}
\put(109,154){\makebox(2,2)[c]{$\scriptstyle N\Delta_1$}}
\put(149,154){\makebox(2,2)[c]{$\scriptstyle N\Delta_2$}}
\put(189,154){\makebox(2,2)[c]{$\scriptstyle N\Delta_3$}}
\put(209,144){\makebox(2,2)[c]{$\scriptstyle a$}}
\put(169,144){\makebox(2,2)[c]{$\scriptstyle b$}}

\put(5,80){\makebox(10,10)[c]{$\displaystyle +$}}

\drawline[0](20,85)(40,85)
\dashline{4}(41,85)(60,85)
\drawline[0](61,85)(80,85)
\dashline{4}(81,85)(100,85)
\dashline{4}(100,65)(100,105)
\drawline[0](100,64)(100,45)
\drawline[0](100,106)(100,125)
\put(30,85){\circle*{3.2}}
\put(70,85){\circle*{3.2}}
\put(100,55){\circle*{3.2}}
\put(100,115){\circle*{3.2}}
\put(21,81){\makebox(2,2)[c]{$\scriptscriptstyle \delta$}}
\put(62,81){\makebox(2,2)[c]{$\scriptscriptstyle \gamma$}}
\put(102,45){\makebox(2,2)[c]{$\scriptscriptstyle \beta$}}
\put(102,123){\makebox(2,2)[c]{$\scriptscriptstyle \alpha$}}
\put(104,114){\makebox(2,2)[c]{$\scriptstyle s$}}
\put(105,55){\makebox(2,2)[c]{$\scriptstyle s'$}}
\put(31,90){\makebox(2,2)[c]{$\scriptstyle s'''$}}
\put(71,90){\makebox(2,2)[c]{$\scriptstyle s''$}}
\put(49,89){\makebox(2,2)[c]{$\scriptstyle N\Delta_1$}}
\put(89,89){\makebox(2,2)[c]{$\scriptstyle N\Delta_2$}}
\put(108,97){\makebox(2,2)[c]{$\scriptstyle N\Delta_4$}}
\put(108,72){\makebox(2,2)[c]{$\scriptstyle N\Delta_3$}}
\put(93,54){\makebox(2,2)[c]{$\scriptstyle b$}}
\put(93,114){\makebox(2,2)[c]{$\scriptstyle a$}}

\put(125,80){\makebox(10,10)[c]{$\displaystyle +$}}

\dashline{4}(160,65)(160,105)
\drawline[0](160,64)(160,45)
\drawline[0](160,106)(160,125)
\dashline{4}(160,85)(230,85)
\dashline{4}(230,65)(230,105)
\drawline[0](230,64)(230,45)
\drawline[0](230,106)(230,125)
\put(160,55){\circle*{3.2}}
\put(160,115){\circle*{3.2}}
\put(230,55){\circle*{3.2}}
\put(230,115){\circle*{3.2}}
\put(232,45){\makebox(2,2)[c]{$\scriptscriptstyle \beta$}}
\put(232,123){\makebox(2,2)[c]{$\scriptscriptstyle \alpha$}}
\put(162,45){\makebox(2,2)[c]{$\scriptscriptstyle \gamma$}}
\put(162,123){\makebox(2,2)[c]{$\scriptscriptstyle \delta$}}
\put(234,114){\makebox(2,2)[c]{$\scriptstyle s$}}
\put(235,55){\makebox(2,2)[c]{$\scriptstyle s'$}}
\put(166,115){\makebox(2,2)[c]{$\scriptstyle s'''$}}
\put(165,55){\makebox(2,2)[c]{$\scriptstyle s''$}}

\put(150,97){\makebox(2,2)[c]{$\scriptstyle N\Delta_1$}}
\put(150,72){\makebox(2,2)[c]{$\scriptstyle N\Delta_2$}}
\put(238,97){\makebox(2,2)[c]{$\scriptstyle N\Delta_4$}}
\put(238,72){\makebox(2,2)[c]{$\scriptstyle N\Delta_3$}}
\put(194,89){\makebox(2,2)[c]{$\scriptstyle N\Delta_5$}}
\put(223,54){\makebox(2,2)[c]{$\scriptstyle b$}}
\put(223,114){\makebox(2,2)[c]{$\scriptstyle a$}}

\end{picture}

\begin{eqnarray}
I)&=&\frac{2}{N}\sum_{p\neq q=1}^{4}\int_{0}^{L_\alpha}ds \int_{0}^{s}ds' \int_{0}^{s'}ds'' \int_{0}^{s''}ds''' \nonumber \\
&&\qquad{\mathcal{P}}\left(s''-s'''+h^{(4)}_{pq}(s'-s'')+s-s' \right) \nonumber \\
&=&8 N^3 \sum_{p=2}^{4}T_1(f_\alpha,h^{(4)}_{1p})  \nonumber 
\end{eqnarray}
\begin{eqnarray}
II)&=&\frac{2}{N}\sum_{p=1,p\neq a}^{4}\int_{0}^{L_\alpha}ds \int_{0}^{L_\beta}ds' \int_{0}^{s'}ds'' \int_{0}^{s''}ds''' \nonumber \\
&&\qquad{\mathcal{P}}\left(s''-s'''+h^{(4)}_{ap}(s'-s'')+L_\beta-s'+N\Delta+s \right)  \nonumber \\
&=&2 N^3 e^{-x \Delta} F_{f_\alpha}\sum_{p=1,p\neq a}^{4}T_2(f_\beta,h^{(4)}_{ap})\nonumber \\
III)&=&\frac{4}{N}\int_{0}^{L_\alpha}ds \int_{0}^{s}ds' \int_{0}^{L_\beta}ds'' \int_{0}^{s''}ds''' \nonumber \\
&&\qquad{\mathcal{P}}\left(s''-s'''+h^{(4)}_{ab}(L_\beta-s''+N\Delta+s')+s-s' \right)  \nonumber \\
&=&4N^3 e^{-h x \Delta} J_{f_\alpha}(h^{(4)}_{ab}) J_{f_\beta}(h^{(4)}_{ab})\nonumber \\
IV)&=&\frac{2}{N}\int_{0}^{L_\alpha}ds \int_{0}^{s}ds' \int_{0}^{L_\beta}ds'' \int_{0}^{L_\gamma}ds''' \nonumber \\
&&\qquad{\mathcal{P}}\left( s'''+N\Delta_1+s''+h^{(4)}_{ab}(L_\beta-s''+N\Delta_2+s')+s-s' \right)  \nonumber \\
&=&2N^3 e^{-x(\Delta_1+h^{(4)}_{ab} \Delta_2)} F_{f_\gamma}e^{-x f_\beta}{\mathcal{F}}_{f_\beta}(h^{(4)}_{ab}-1)J_{f_\alpha}(h^{(4)}_{ab})\nonumber \\
V)&=&\frac{1}{N}\sum_{p=1,p\neq a,b}^{4}\int_{0}^{L_\alpha}ds \int_{0}^{L_\beta}ds' \int_{0}^{s'}ds'' \int_{0}^{L_\gamma}ds''' \nonumber \\
&&\qquad{\mathcal{P}}\left(s'''+N\Delta_1+s''+h^{(4)}_{ap}(s'-s'')+L_\beta-s'+N\Delta_2+s \right)  \nonumber \\
&=&N^3 e^{-x(\Delta_1+\Delta_2)}F_{f_\alpha}F_{f_\gamma}e^{-f_\beta x}\sum_{p=1,p\neq a,b}^{4} {\mathcal{D}}_{f_\beta}(h^{(4)}_{ap}-1)\nonumber \\
VI)&=&\frac{2}{N}\int_{0}^{L_\alpha}ds \int_{0}^{L_\beta}ds' \int_{0}^{L_\gamma}ds'' \int_{0}^{s''}ds''' \nonumber \\
&&\qquad{\mathcal{P}}\left( 
s+N\Delta_1+s'+N\Delta_2+h^{(4)}_{ab}(N\Delta_3+s''')+s''-s''' \right)  \nonumber \\
&=&2N^3 e^{-x(\Delta_1+\Delta_2+h^{(4)}_{ab}\Delta_3)}F_{f_\alpha}F_{f_\beta}J_{f_\gamma}(h^{(4)}_{ab})\nonumber \\
VII)&=&\frac{1}{N}\int_{0}^{L_\alpha}ds \int_{0}^{L_\beta}ds' \int_{0}^{L_\gamma}ds'' \int_{0}^{L_\delta}ds''' \nonumber \\
&&\qquad{\mathcal{P}}\left(s'''+N\Delta_1+s''+h^{(4)}_{ab}(L_\gamma-s''+N\Delta_2+s')+L_\beta-s'+N\Delta_3+s \right)  \nonumber \\
&=&N^3 e^{-x(\Delta_1+h^{(4)}_{ab}\Delta_2+\Delta_3)}F_{f_\alpha}{\mathcal{F}}_{f_\beta}(h^{(4)}_{ab}-1)e^{-x(f_\beta+f_\gamma)} {\mathcal{F}}_{f_\gamma}(h^{(4)}_{ab}-1) F_{f_\delta}\nonumber \\
VIII)&=&\frac{1}{N}\int_{0}^{L_\alpha}ds \int_{0}^{L_\beta}ds' \int_{0}^{L_\gamma}ds'' \int_{0}^{L_\delta}ds''' \nonumber \\
&&\qquad{\mathcal{P}}\left(
s+N\Delta_4 + s'+N\Delta_3 +s'''+N\Delta_1+s''+h^{(4)}_{ab}(L_\gamma-s''+N\Delta_2) \right)  \nonumber \\
&=&N^3 e^{-x(\Delta_1+h^{(4)}_{ab}\Delta_2+\Delta_3+\Delta_4)}F_{f_\alpha}F_{f_\beta}e^{f_\gamma x}{\mathcal{F}}_{f_\gamma}(h^{(4)}_{ab}-1)F_{f_\delta}\nonumber 
\end{eqnarray}
\begin{eqnarray}
IX)&=&\frac{1}{N}\int_{0}^{L_\alpha}ds \int_{0}^{L_\beta}ds' \int_{0}^{L_\gamma}ds'' \int_{0}^{L_\delta}ds''' \nonumber \\
&&\qquad{\mathcal{P}}\left(
s+N\Delta_4+s'+N\Delta_3+h^{(4)}_{ab}N\Delta_5+s''+N\Delta_2+s'''+N\Delta_1 \right)  \nonumber \\
&=&N^3 F_{f_\alpha} F_{f_\beta} F_{f_\gamma} F_{f_\delta} e^{-x(\Delta_1+\Delta_2+\Delta_3+\Delta_4+h^{(4)}_{ab}\Delta_5)} \nonumber
\end{eqnarray}

\section{Coefficients $\Xi$}

We use the following notation: $\displaystyle\tilde{\Gamma}_4\left(\sum_{\alpha=1}^{4}\frac{\Delta_\alpha}{q_*},h_1,h_2\right)=\tilde{\Gamma}_4\left(\sum_{\alpha=1}^{4}\frac{\Delta_\alpha}{q_*},h^{(4)}\right)$.

FCC:
\begin{eqnarray}
\Xi_1^{(F)}&=&6\tilde{\Gamma}_4\left(0,0,\frac{4}{3}\right)+\tilde{\Gamma}_4\left(0,0,0\right)+2\tilde{\Gamma}_4\left(0,\frac{4}{3},\frac{4}{3}\right) \nonumber \\
\Xi_2^{(F)}&=&4\tilde{\Gamma}_4\left(4Q,0,2\right)+\tilde{\Gamma}_4\left(4Q,0,0\right) \nonumber \\
\Xi_3^{(F)}&=&3\tilde{\Gamma}_4\left(2Q,0,z\right)+\tilde{\Gamma}_4\left(2Q,2,z\right)+\tilde{\Gamma}_4\left(2Q,z,z\right)+\tilde{\Gamma}_4\left(2Q,\frac{8}{3},z\right) \nonumber \\
&&\quad z=2-\frac{2}{\sqrt{3}} \qquad,\qquad Q=\frac{2}{\sqrt{3}}-1 \nonumber
\end{eqnarray}

OBDD:
\begin{eqnarray}
\Xi_1^{(O)}&=&2\tilde{\Gamma}_4\left(0,0,2\right)+8\tilde{\Gamma}_4\left(0,0,1\right)+\tilde{\Gamma}_4\left(0,0,0\right)+4\tilde{\Gamma}_4\left(0,1,1\right) \nonumber \\
\Xi_2^{(O)}&=&\tilde{\Gamma}_4\left(4Q,0,0\right)+2\tilde{\Gamma}_4\left(4Q,\frac{4}{3},\frac{4}{3}\right)+6\tilde{\Gamma}_4\left(4Q,0,\frac{4}{3}\right) \nonumber \\
\Xi_3^{(O)}&=&4\tilde{\Gamma}_4\left(2Q,0,2\right)+6\tilde{\Gamma}_4\left(2Q,0,z\right)-2\tilde{\Gamma}_4\left(2Q,2,\frac{2}{3}\right)-6\tilde{\Gamma}_4\left(2Q,2,z\right)+\tilde{\Gamma}_4\left(2Q,z,z\right) \nonumber \\
&& \quad+\tilde{\Gamma}_4\left(2Q,4,z\right)+\tilde{\Gamma}_4\left(2Q,\frac{8}{3},z\right)-2\tilde{\Gamma}_4\left(2Q,\frac{4}{3},z\right) \nonumber \\
&&\quad z=2-2\sqrt{\frac{2}{3}}\qquad,\qquad Q=\sqrt{\frac{3}{2}}-1 \nonumber
\end{eqnarray}

GYR:
\begin{eqnarray}
\Xi_1^{(G)}&=&\tilde{\Gamma}_4\left(0,0,0\right)+4\tilde{\Gamma}_4\left(0,0,\frac{1}{3}\right)+4\tilde{\Gamma}_4\left(0,0,1\right)+8\tilde{\Gamma}_4\left(0,0,\frac{5}{3}\right)-4\tilde{\Gamma}_4\left(0,\frac{1}{3},\frac{2}{3}\right) \nonumber \\
&-&4\tilde{\Gamma}_4\left(0,\frac{5}{3},\frac{5}{3}\right)+2\tilde{\Gamma}_4\left(0,0,\frac{4}{3}\right)+2\tilde{\Gamma}_4\left(0,\frac{2}{3},\frac{2}{3}\right)+4\tilde{\Gamma}_4\left(0,0,\frac{2}{3}\right) \nonumber \\ \nonumber \\
\Xi_2^{(G)}&=&8\tilde{\Gamma}_4\left(4Q,0,1\right)+2\tilde{\Gamma}_4\left(4Q,0,2\right)+4\tilde{\Gamma}_4\left(4Q,1,1\right)+\tilde{\Gamma}_4\left(4Q,0,0\right)\nonumber \\ \nonumber
\end{eqnarray}
\begin{eqnarray}
\Xi_3^{(G)}&=&2\tilde{\Gamma}_4\left(Q,0,\frac{1}{3}\right)-2\tilde{\Gamma}_4\left(Q,0,\frac{5}{3}\right)+\tilde{\Gamma}_4\left(Q,\frac{1}{3},\frac{1}{3}\right)+\tilde{\Gamma}_4\left(Q,\frac{2}{3},\frac{5}{3}\right)+2\tilde{\Gamma}_4\left(Q,1,\frac{4}{3}\right) \nonumber \\
&-&6\tilde{\Gamma}_4\left(Q,2,\frac{1}{3}\right)-2\tilde{\Gamma}_4\left(Q,2,\frac{1}{\sqrt{3}}\right)-2\tilde{\Gamma}_4\left(Q,2,\sqrt{3}\right)-2\tilde{\Gamma}_4\left(Q,\frac{1}{3},\frac{4}{3}\right)\nonumber \\
&+&2\tilde{\Gamma}_4\left(Q,\frac{5}{3},\frac{1}{3}+\frac{2}{\sqrt{3}}\right) -2\tilde{\Gamma}_4\left(Q,\frac{1}{3},\frac{5}{3}+\frac{1}{\sqrt{3}}\right)
+2\tilde{\Gamma}_4\left(Q,1,1+\frac{1}{\sqrt{3}}\right) \nonumber \\
&+&2\tilde{\Gamma}_4\left(Q,\frac{5}{3},\frac{1}{3}+\frac{1}{\sqrt{3}}\right)-2\tilde{\Gamma}_4\left(Q,\frac{7}{3},\frac{1}{\sqrt{3}}-\frac{1}{3}\right) 
+2\tilde{\Gamma}_4\left(Q,\sqrt{3},2-\frac{1}{\sqrt{3}}\right)\nonumber \\
&+&\tilde{\Gamma}_4\left(Q,\frac{2}{\sqrt{3}},2-\frac{1}{\sqrt{3}}\right)+2\tilde{\Gamma}_4\left(Q,\frac{1}{3},\frac{5}{3}-\frac{2}{\sqrt{3}}\right)+2\tilde{\Gamma}_4\left(Q,\frac{1}{3},\frac{5}{3}+\sqrt{3}\right)\nonumber \\
&-&2\tilde{\Gamma}_4\left(Q,\frac{5}{3},\frac{1}{3}+\sqrt{3}\right) -2\tilde{\Gamma}_4\left(Q,\frac{7}{3},\sqrt{3}-\frac{1}{3}\right)+2\tilde{\Gamma}_4\left(Q,\frac{11}{3},2-\sqrt{3}\right)\nonumber \\
&-&2\tilde{\Gamma}_4\left(Q,2-\sqrt{3},\frac{4}{\sqrt{3}}\right)+2\tilde{\Gamma}_4\left(Q,\frac{1}{\sqrt{3}},2-\sqrt{3}\right) +\tilde{\Gamma}_4\left(Q,2\sqrt{3},2-\sqrt{3}\right) \nonumber \\ \nonumber
 \\
\Xi_4^{(G)}&=&6\tilde{\Gamma}_4\left(2Q,0,2\right)-2\tilde{\Gamma}_4\left(2Q,2,\frac{1}{\sqrt{3}}\right)-2\tilde{\Gamma}_4\left(2Q,2,\sqrt{3}\right)-2\tilde{\Gamma}_4\left(2Q,\frac{8}{3},2-\frac{1}{\sqrt{3}}\right) \nonumber \\
&-&2\tilde{\Gamma}_4\left(2Q,\frac{8}{3},2-\sqrt{3}\right)-2\tilde{\Gamma}_4\left(2Q,1,2+\frac{1}{\sqrt{3}}\right)-2\tilde{\Gamma}_4\left(2Q,\frac{4}{3},2+\frac{1}{\sqrt{3}}\right)\nonumber \\
&+&12\tilde{\Gamma}_4\left(2Q,0,2+\frac{1}{\sqrt{3}}\right)+6\tilde{\Gamma}_4\left(2Q,0,2+\frac{2}{\sqrt{3}}\right)-2\tilde{\Gamma}_4\left(2Q,1,1+\sqrt{3}\right)
\nonumber \\
&-&2\tilde{\Gamma}_4\left(2Q,\frac{4}{3},\frac{2}{3}+\sqrt{3}\right)+2\tilde{\Gamma}_4\left(2Q,3,\sqrt{3}-1\right)+2\tilde{\Gamma}_4\left(2Q,4,2-\sqrt{3}\right)\nonumber \\
&-&4\tilde{\Gamma}_4\left(2Q,2-\sqrt{3},\frac{4}{\sqrt{3}}\right)-4\tilde{\Gamma}_4\left(2Q,2-\sqrt{3},\frac{2}{\sqrt{3}}\right)+2\tilde{\Gamma}_4\left(2Q,2-\sqrt{3},2\sqrt{3}\right) \nonumber \\
&+&12\tilde{\Gamma}_4\left(2Q,0,2-\sqrt{3}\right) \nonumber \\
&& \qquad \qquad\qquad \qquad \qquad \qquad Q=\frac{2}{\sqrt{3}}-1 \nonumber
\end{eqnarray}

\bibliographystyle{unsrt}
\bibliography{my}

\end{document}